\documentclass[10pt,a4paper]{article}

\usepackage{caption}
\usepackage{subcaption}
\usepackage{graphicx}
\graphicspath{{figures/}}

\usepackage{amssymb}
\usepackage{amsmath} 

\usepackage{array}
\renewcommand{\arraystretch}{1.3} 

\usepackage{algorithmic}
\usepackage{array}
\usepackage{lipsum}
\usepackage{hyperref}
\usepackage{float}
\usepackage{booktabs}
\usepackage{dcolumn}
\usepackage{multirow}
\usepackage{makecell}
\usepackage{bigints}
\usepackage{xspace}

\begin{document}

\title{\Large Development of a Cost-Effective Simulation Tool for Loss of Flow Accident Transients in High-Temperature Gas-cooled Reactors} 
\author{\normalsize \textbf{Bo Liu$^{1,*}$, Wei Wang$^1$, Charles Moulinec$^1$, Stefano Rolfo$^1$, Marion Samler$^1$,} \\ \normalsize \textbf{Ehimen Iyamabo$^2$, Constantinos Katsamis$^2$ and Marc Chevalier$^2$} \\ \normalsize $^1$ Scientific Computing, Science and Technology Facilities Council \\ \normalsize $^2$ EDF Energy R\&D UK Centre \\ \normalsize $^*$\href{mailto:bo.liu@stfc.ac.uk}{bo.liu@stfc.ac.uk}}
\date{7 March, 2025} 
\maketitle 

\begin{abstract}
\normalsize
The aim of this work is to further expand the capability of the coarse-grid Computational Fluid Dynamics (CFD) approach, SubChCFD, to effectively simulate transient and buoyancy-influenced flows, which are critical in accident analyses of High-Temperature Gas-cooled Reactors (HTGRs). It has been demonstrated in our previous work that SubChCFD is highly adaptable to HTGR fuel designs and performs exceptionally well in modelling steady-state processes. In this study, the approach is extended to simulate a Loss of Flow Accident (LOFA) transient, where coolant circulation is disrupted, causing the transition from forced convection to buoyancy-driven natural circulation within the reactor core. To enable SubChCFD to capture the complex physics involved, corrections were introduced to the empirical correlations to account for the effects of flow unsteadiness, property variation and buoyancy.
\smallskip

A $1/12$th sector of the reactor core, representing the smallest symmetric unit, was modelled using a coarse mesh of approximately $60$ million cells. This mesh size is about $6\%$ of that required for a Reynolds Averaged Navier Stokes (RANS) model, where mesh sizes can typically reach the order of $1$ billion cells for such configurations. Simulation results show that SubChCFD effectively captures the thermal hydraulic behaviours of the reactor during a LOFA transient, producing predictions in good agreement with RANS simulations while significantly reducing computational cost.
\smallskip

\noindent
\textbf{Key words:} Computational Fluid Dynamics; Coarse-grid; Subchannel; High Temperature Gas-cooled Reactor; Loss of Flow Accident

\end{abstract}
\clearpage
\setcounter{tocdepth}{2}
\tableofcontents

\clearpage

\section{Introduction}
\label{sec:introduction}

As a low-carbon energy source, nuclear power provides a reliable and consistent supply of electricity without directly producing greenhouse gas emissions. As part of the UK government's strategy to achieve Net-Zero, nuclear energy is expected to contribute 24 GW to the electricity grid by 2050. Advanced nuclear technologies, particularly the Generation IV reactors \cite{Locatelli2013}, offer improved safety features, higher efficiency, and reduced waste generation. By embracing these advanced techniques, the UK can enhance the sustainability and efficiency of its nuclear power infrastructure.
\smallskip

Leveraging the nation's extensive experience in operating gas-cooled reactors, such as Advanced Gas-cooled Reactors (AGRs), the UK government is actively pursuing the development of High-Temperature Gas-cooled Reactor (HTGR) technology. HTGRs are widely recognised for their inherent safety features, high thermal efficiency, and ability to provide high-temperature process heat, making them a promising candidate for next-generation nuclear reactors. Given these advantages, the current ambition in the UK is to deploy this technology, aiming to have a demonstrator operational by the early 2030s \cite{BEIS2022}.
\smallskip

HTGRs are designed with passive safety features to mitigate the consequences of extreme operating conditions or accident scenarios. One such scenario is the Loss of Forced Flow Accident (LOFA), in which forced convection cooling is lost, and heat removal must rely on a combination of natural convection, radiation, and conduction. Accurately modelling the complex thermal-hydraulic behaviour during LOFA is crucial for reactor safety analysis and design optimisation. Computational Fluid Dynamics (CFD), which can accurately predict the 3-D temperature distribution within the reactor core, is an indispensable tool for such modelling, providing support to the safety claims of HTGRs under LOFA conditions. For instance, Tung et al. \cite{Tung2016} conducted CFD simulations to investigate the thermal-hydraulic response of an HTGR core, using a series of models of varying scales. These include a 1/12th sector model (only a fraction of a single fuel assembly), a two-sub-region model (covering 24 symmetric 1/12th sectors), and a 1/12th core model (the smallest symmetric representation of the entire core). Their findings indicated that while smaller models provided reasonable accuracy for steady-state conditions, they were insufficient for simulating LOFA transients due to excessive simplifications. The two-sub-region model performed significantly better than the 1/12th sector model but still failed to adequately capture the transient process. The most detailed and accurate results were only obtained using the 1/12th core model, highlighting the necessity of large-scale (core-scale) models to accurately predict the flow behaviour and temperature distribution across the entire reactor core. 
\smallskip

Unfortunately, large scale CFD simulations are computationally expensive. Despite advancements in High-Performance Computing (HPC) technology in the past decades, full-scale CFD simulations of a reactor core remain impractical, making them unfeasible for routine engineering design calculations and safety assessments. For example, in Tung's work, only 100 seconds of the LOFA were simulated, a very small fraction of the full accident timeline, due to the high demand in computational resources. This is insufficient for accurately capturing the transition from forced convection to fully established natural circulation during a LOFA event, as this transition typically occurs over a timescale of tens of minutes to hours. In such a context, there is an urgent need to have an efficient, cost-effective, and accurate thermal hydraulics modelling tool that retains full 3-D capabilities while significantly reducing computational overhead.
\smallskip

Coarse-grid CFD is a promising 3-D methodology that combines the advantages of established CFD while keeping the computational cost reasonable when applied to full-scale engineering geometries. In previous work, a coarse-grid CFD approach known as Subchannel CFD (SubChCFD), implemented based on the open source CFD software $code_saturne$ (v8.0), was successfully extended to prismatic HTGR fuel assemblies. This method demonstrated the ability to provide 3-D flow and thermal field predictions with significantly reduced computational costs compared to conventional Reynolds-Averaged Navier-Stokes (RANS) approaches \cite{Liu2019}. However, the initial development of SubChCFD primarily focused on steady-state operating conditions in prismatic HTGRs. Building on this foundation, the current project aims to extend the capabilities of SubChCFD to accurately model HTGRs under LOFA conditions. This extension involves addressing several critical challenges associated with transient thermal-hydraulic phenomena:
\smallskip

\begin{itemize}
    \item \textbf{Transient friction modelling:} A transient friction model is introduced to account for the time-dependent nature of wall shear stress variations during LOFA, ensuring an accurate representation of flow resistance due to inertia of accelerating or decelerating flows.
    \item \textbf{Variable property correction:} Advanced corrections to the friction factor and heat transfer models are implemented to account for the effects of fluid property variation.
    \item \textbf{Buoyancy correction:} Corrections are introduced to incorporate the buoyancy effects, which dominate heat transfer of the coolant in the absence of forced convection.
\end{itemize}
\smallskip

By integrating these technical advancements, the enhanced SubChCFD framework provides a cost-effective solution for modelling HTGR thermal hydraulics under transient conditions. This development is crucial for reactor safety analysis, supporting the design and validation of passive cooling strategies to ensure the long-term stability and integrity of HTGR cores during LOFA events.
\smallskip

The following sections of this report detail the methodological development, numerical implementation, and validation of the extended SubChCFD model, along with a case study demonstrating its applicability to prismatic HTGR fuel assemblies under LOFA conditions.
\section{Development of new functionalities in SubChCFD}
\label{sec:development}

\subsection{Flow unsteadiness}
\label{sec:development:subsection1}
The core principle of SubChCFD is to use empirical friction and heat transfer correlations to account for unresolved near-wall effects, allowing for the use of coarse computational meshes. However, these correlations are primarily validated for steady-state flows and can lead to significant inaccuracies in transient scenarios, particularly under LOFA conditions. During a LOFA, events such as a pump trip can cause a rapid slowdown in flow, followed by flow acceleration as natural circulation is established. In such cases, wall shear stress varies significantly over time, which cannot be accurately captured by conventional steady-state friction models. This transient variation in wall shear stress can directly influence flow redistribution among subchannels, thereby affecting the overall thermal hydraulic behaviour of the reactor core. To improve SubChCFD for LOFA analysis, it is crucial to introduce corrections to the existing friction models, improving the prediction the dynamics of unsteady\footnote{In CFD, unsteady flow refers to any time-dependent flow, including periodic or chaotic behaviours such as turbulence. Transient flow is a specific type of unsteady flow characterized by a system transitioning between states, such as during startup or shutdown processes. While all transient flows are unsteady, not all unsteady flows are transient. In LOFA scenarios, the flow is inherently transient. Therefore, both “unsteady” and “transient” are used interchangeably in this report where appropriate.} wall shear stress.
\smallskip

Research on transient friction models dates back to the 1950s, with early work by Daily et~al. \cite{Daily1956}, who observed that the quasi-steady friction approximation produced an insufficient amount of damping in transient pipe flows as compared to experimental behaviours. To address this, various models were proposed over the following decades, among which the most notable ones are summarised below:
\begin{itemize}
    \item \textbf{Zielke's model \cite{Zielke1968}:} This model accurately captures unsteady friction in laminar flows by incorporating the full velocity history. It was later extended by Vardy et al. \cite{VardyBrown2003} to account for turbulent flows up to a Reynolds number of $10^8$. While effective, its application is limited due to complex formulation and high demand in computational storage.
    \item \textbf{Brunone’s model \cite{VardyBrown1995}:} In contrast to Zielke's model, the Brunone model offers a more computationally efficient alternative by directly relating transient friction to instantaneous acceleration. The key model parameter, $k_3$, was originally determined through experiments but was later deduced with theoretically derived values \cite{VardyBrown2003}. This enhancement significantly improved the model’s applicability, making it widely used in engineering applications such as water hammer studies. \end{itemize}
\smallskip

More recently, new understandings of the dynamics of transient flow have been developed with the advancement of new experimental techniques and high-fidelity CFD simulations. Unsteady turbulent flows exhibit distinct stages of development. Vardy et~al. \cite{VardyBrown2003} first pointed out that unsteady wall friction is influenced by two factors, (i) inertia due to flow acceleration or deceleration, and (ii) delayed response in production of turbulence. He et~al. \cite{HeSeddighi2015} demonstrated that when Reynolds number changes abruptly in transient channel flows, turbulence does not immediately adapt but instead undergoes a laminar-turbulent bypass transition. Mathur et~al. \cite{Mathur2018} examined temporal acceleration in turbulent channel flows and found that initial turbulence does not solely govern transient flow evolution. Instead, inertia and boundary layer instabilities significantly influence the transition process. When the flow is suddenly accelerated, turbulence production lags, leading to an initial period where turbulence intensity remains unchanged despite the increased bulk velocity \cite{Gorji2014}. Consequently, the initial wall shear stress overshoots the quasi-steady value due to inertial effects, followed by an undershoot caused by delayed turbulence response \cite{HeAriyaratne2008}.
\smallskip

\begin{figure}[ht]
    \centering\includegraphics[width=0.6\linewidth]{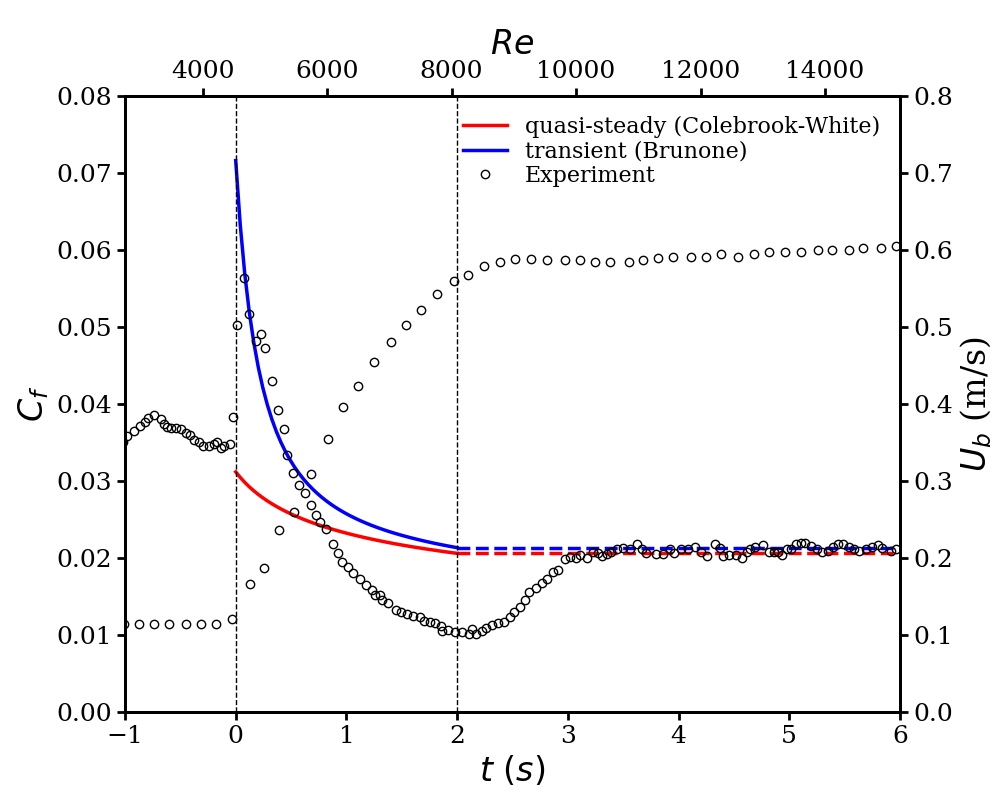}
    \caption{Benefits gain by using the Brunone model in capturing the inertia effect in accelerating flows. The data used here is based on a typical case in Mathur et al. \cite{Mathur2018}, with flow accelerating from Reynolds number of 2,760 to 15,200 in a plane channel.}
    \label{fig:Brunone_cf}
\end{figure}

Despite the progress in understanding transient turbulent flows, developing friction models that fully capture the associated physics remains a significant challenge due to the complexity of the problem. Among the available models, the Brunone model appears to be an ideal choice for integration into SubChCFD, as it provides a good balance between accuracy and computational efficiency for full-scale reactor simulations. This is because the model relies solely on instantaneous acceleration, no historical data is required. As shown in Figure \ref{fig:Brunone_cf}, the Brunone model effectively accounts for the immediate response of wall shear stress to acceleration by incorporating local inertia corrections, making it an improvement over traditional quasi-steady friction models for accelerating flows. However, it does not capture the delayed effects of turbulence response.
\smallskip

The implementation of the Brunone model in SubChCFD is straightforward. This model directly relates the unsteady component of friction to the instantaneous local acceleration of the bulk flow. The mathematical expression of the model is written as follows,

\begin{equation} \label{eq:brunone}
    C_f = {C_{fs}} + \frac{{k_3D}}{{U|U|}}\left( {\frac{{\partial U}}{{\partial t}} - a\frac{{\partial U}}{{\partial x}}} \right)
\end{equation}

\noindent
where $C_{fs}$ is the steady component of the friction factor, calculated in SubChCFD using the following correlation,

\begin{equation} \label{eq:Cfs}
C_{fs} =
\begin{cases} 
     \text{Laminar flow:} & \quad \frac{64}{\text{Re}} \\[10pt]
    \text{Turbulent flow:} & \quad \frac{1}{\sqrt{C_{fs}}} = -2.0 \log \left( \frac{2.51}{\text{Re} \sqrt{C_{fs}}} \right)
\end{cases}
\end{equation}

\noindent
$U$ represents the local mean velocity, while $k_3$ denotes Brunone's friction coefficient, which can be derived analytically or through experimental trials. Vardy et al. \cite{VardyBrown2003} claimed that the $k_3$ should not be treated as a constant but rather as a function of the local Reynolds number. Based on their shear decay theory, $k_3$ can be derived analytically and expressed as $k_3=\sqrt{C^*}/2$, where $C^*$ is the shear decay coefficient,

\begin{equation} \label{eq:Cstar}
    C^* =
    \begin{cases} 
        \text{Laminar flow:} \quad 0.00476 \\[8pt]
        \text{Turbulent flow:} \quad \frac{12.86}{\text{Re}^\kappa}
    \end{cases}
\end{equation}

\noindent
with,

\begin{equation} \label{eq:kappa}
    \kappa  = {\log _{10}}\left( {\frac{{15.29}}{{{{{\mathop{\rm Re}\nolimits} }^{0.0567}}}}} \right)
\end{equation}

To assess the performance of the newly implemented friction model in SubChCFD, test cases are designed based on transient conditions relevant to LOFA scenarios in an HTGR core. Temporal acceleration (ramp-up) and deceleration (ramp-down) flows are considered in a circular pipe with a diameter of $D=15.88$ mm, the same as that of the coolant channels of a typical HTGR core. Helium is used as work fluid. Three reference temperatures, 500 $^\circ C$, 750 $^\circ C$ and 950 $^\circ C$, are selected to represent the states of the coolant at locations of inlet, middle and outlet of the HTGR core, respectively. In ramp-up flows, the initial velocity is fixed at 10 m/s, while the final velocity is estimated based on steady-state operating conditions at the respective reference temperatures. Conversely, in ramp-down flows, the final velocity is set at 10 m/s, with the initial velocity estimated based on the reactor's operating conditions. Three transient intensities are considered, characterised by transient durations of $t=1.0 s$, $t=0.1 s$ and $t=0.01 s$. More details of these test cases can be found in Table \ref{tab:ramp_up_down}.

\floatstyle{plaintop}
\restylefloat{table}
\begin{table}[h]
    \centering
    \renewcommand{\arraystretch}{1.0}
    \setlength{\tabcolsep}{8pt} 
    \begin{tabular}{c | c | c c | c c}
        \hline\hline
        \textbf{Reference} & \textbf{$t$ (s)} & \multicolumn{2}{c|}{\textbf{Rump-up}} & \multicolumn{2}{c}{\textbf{Rump-down}} \\
        \textbf{Temperature (°C)} & & $U_0$ (m/s) & $U_1$ (m/s) & $U_0$ (m/s) & $U_1$ (m/s) \\
        \hline
        & 1.0  & 10 & 35 & 35 & 10 \\
        500 & 0.1  & 10 & 35 & 35 & 10 \\
        & 0.01 & 10 & 35 & 35 & 10 \\
        & 1.0  & 10 & 45 & 45 & 10 \\
        750 & 0.1  & 10 & 45 & 45 & 10 \\
        & 0.01 & 10 & 45 & 45 & 10 \\
        & 1.0  & 10 & 55 & 55 & 10 \\
        950 & 0.1  & 10 & 55 & 55 & 10 \\
        & 0.01 & 10 & 55 & 55 & 10 \\
        \hline\hline
    \end{tabular}
    \caption{Test cases of ramp-up and ramp-down flows at different reference temperatures and acceleration/deceleration intensities.}
    \label{tab:ramp_up_down}
\end{table}

\begin{figure}[ht]
    \centering\includegraphics[width=0.8\linewidth]{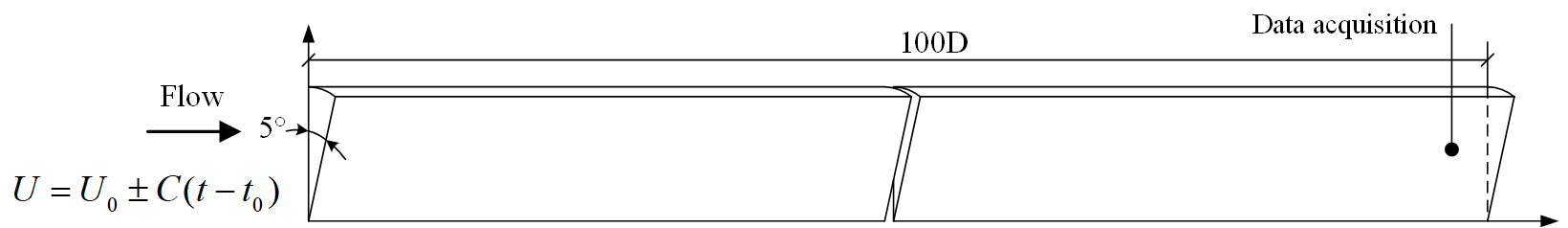}
    \caption{5-degree wedge for the RANS simulations.}
    \label{fig:2d_pipe}
\end{figure}

RANS models are employed to produce reference results for these test cases. The simulations are conducted using a 5-degree wedge domain with a length of $100D$, as illustrated in Figure \ref{fig:2d_pipe}. The low Reynolds number Launder-Sharma $k-\epsilon$ model is selected to account for turbulence, which was found to be one of the best models that was able to produce very closely the features of turbulence dynamics exhibited in transient flow experiments \cite{HeAriyaratne2008}. 

\begin{figure}[H]
    \centering
    \begin{subfigure}[t]{0.45\textwidth}
        \centering\includegraphics[width=1\linewidth]{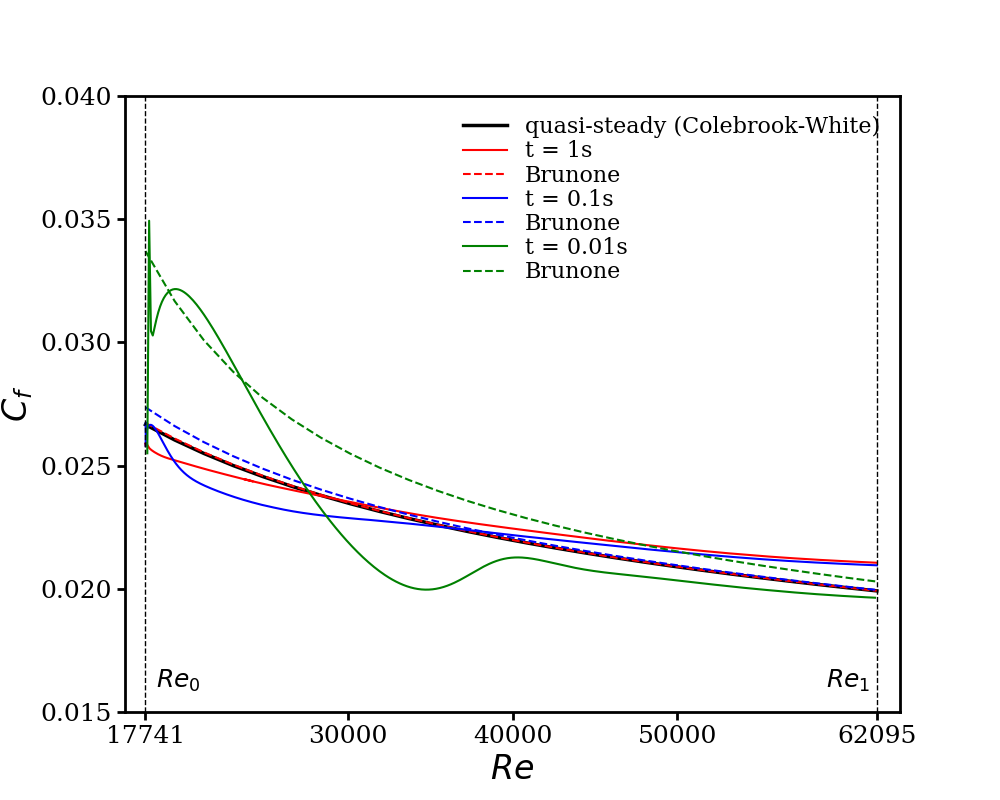}
        \caption{Ramp-up transient at 500 $^\circ C$}
    \end{subfigure}
    \begin{subfigure}[t]{0.45\textwidth}
        \centering\includegraphics[width=1\linewidth]{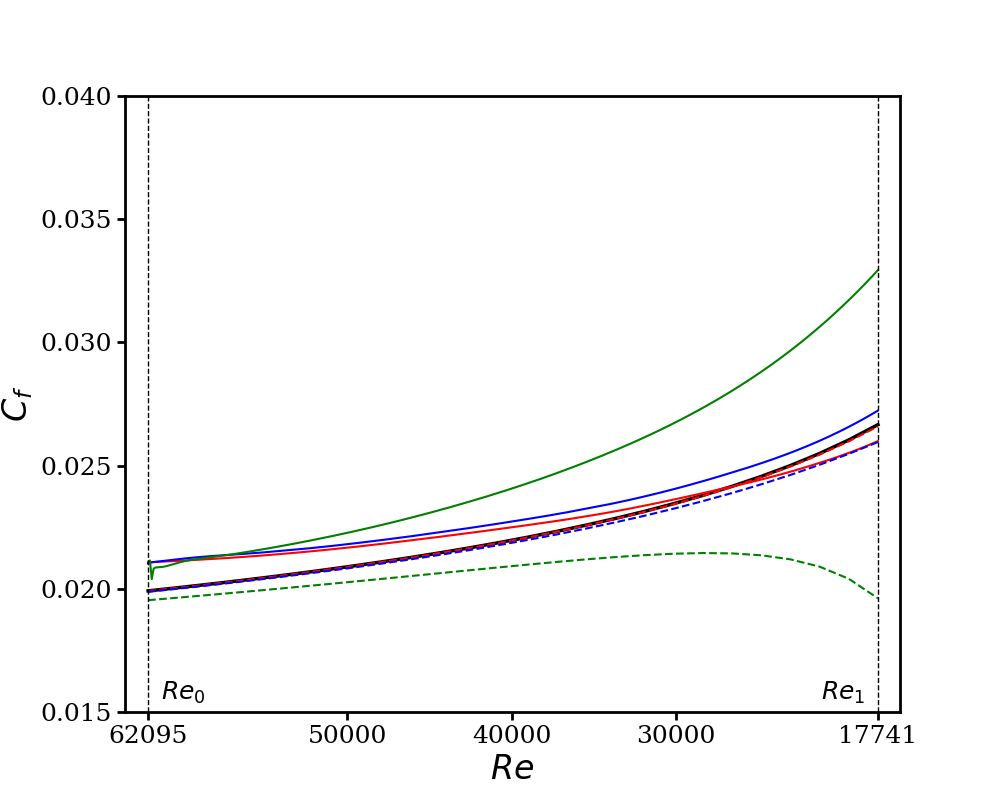}
        \caption{Ramp-down transient at 500 $^\circ C$}
    \end{subfigure}
    \begin{subfigure}[t]{0.45\textwidth}
        \centering\includegraphics[width=1\linewidth]{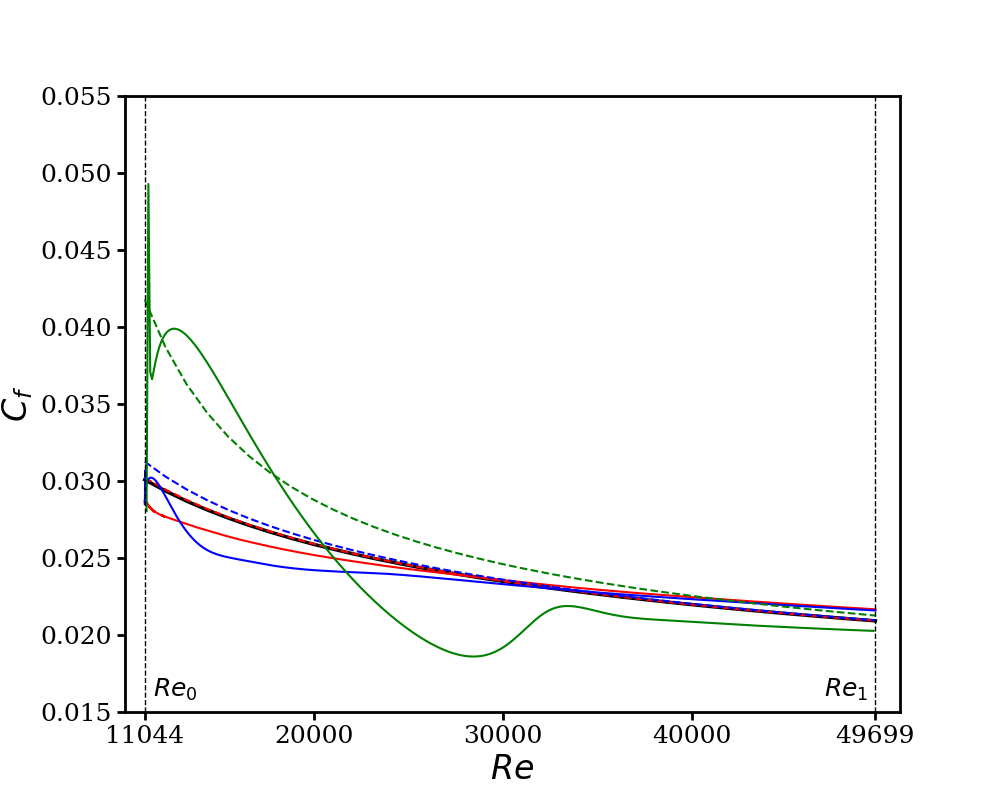}
        \caption{Ramp-up transient at 750 $^\circ C$}
    \end{subfigure}
    \begin{subfigure}[t]{0.45\textwidth}
        \centering\includegraphics[width=1\linewidth]{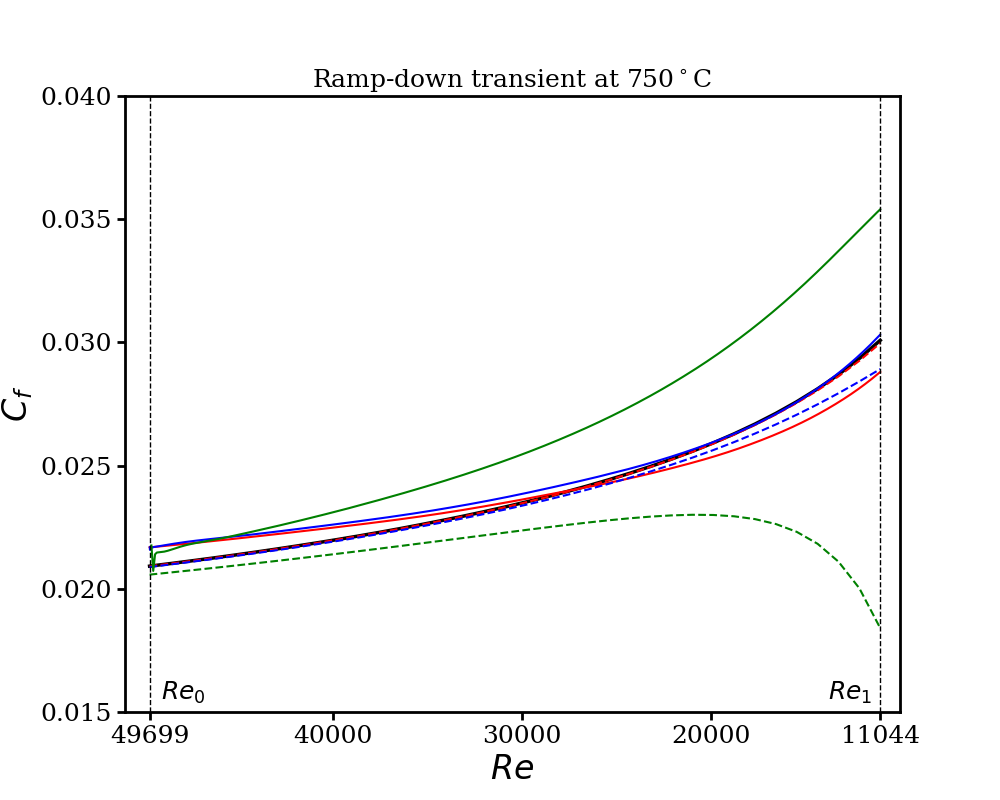}
        \caption{Ramp-down transient at 750 $^\circ C$}
    \end{subfigure}
    \begin{subfigure}[t]{0.45\textwidth}
        \centering\includegraphics[width=1\linewidth]{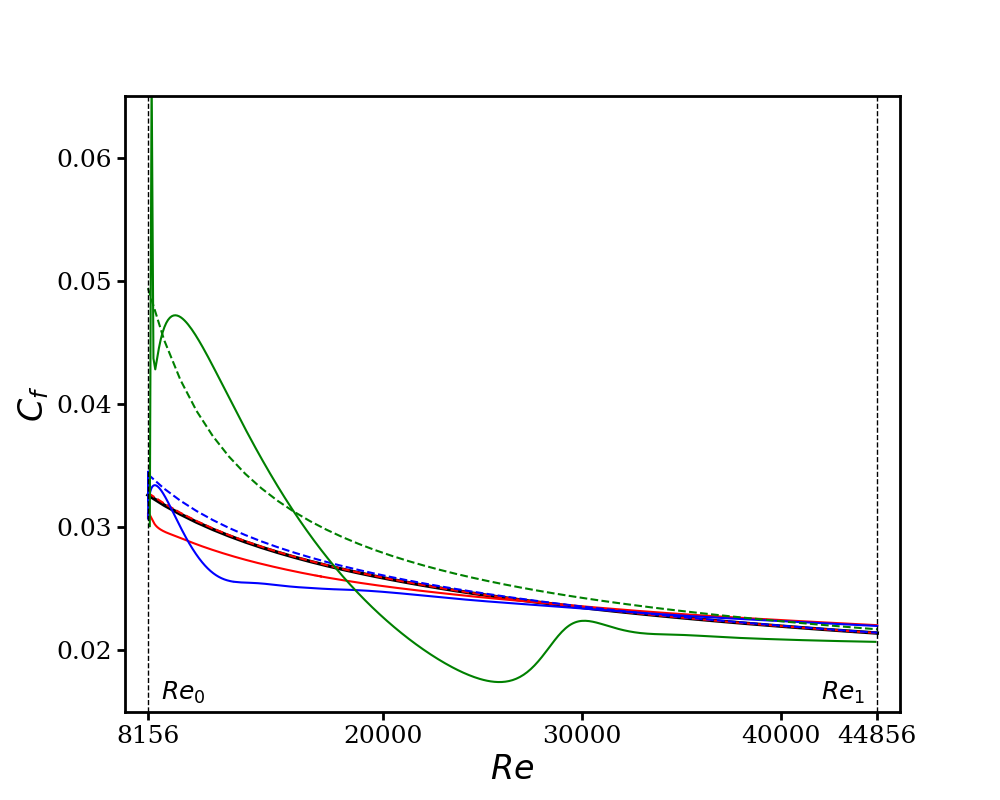}
        \caption{Ramp-up transient at 950 $^\circ C$}
    \end{subfigure}
    \begin{subfigure}[t]{0.45\textwidth}
        \centering\includegraphics[width=1\linewidth]{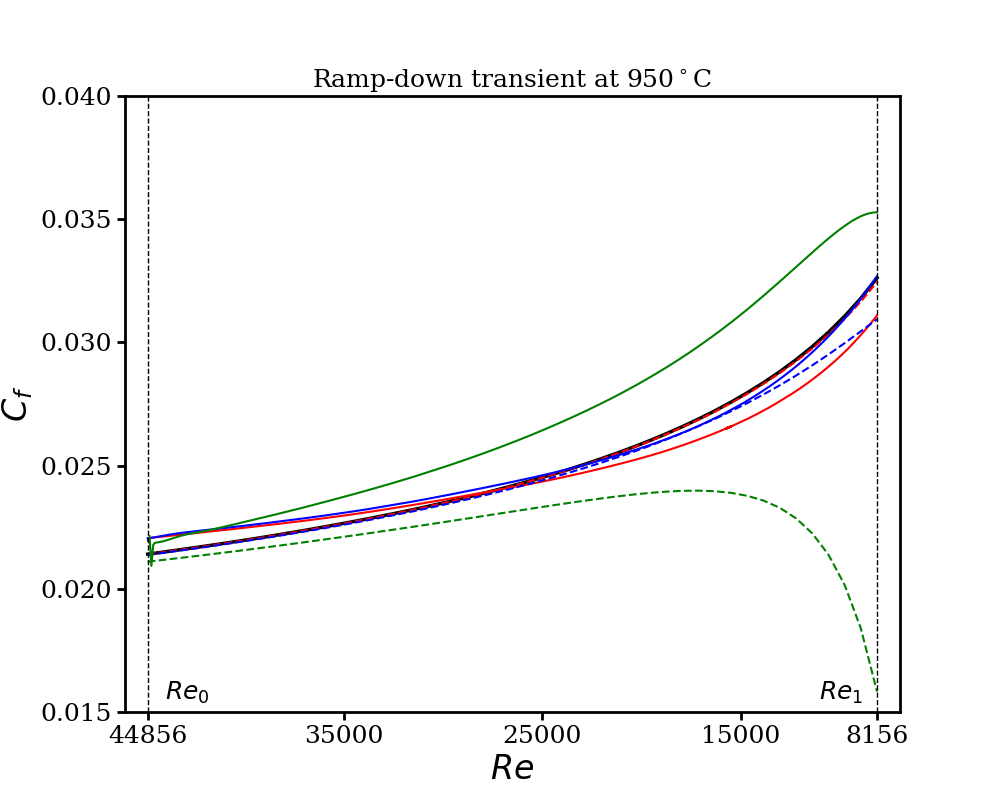}
        \caption{Ramp-down transient at 950 $^\circ C$}
    \end{subfigure}
    \caption{Predictions of $C_f$ in transient ramp-up and ramp-down flows at various reference temperatures and transient intensities.}
    \label{fig:transient_cf}
\end{figure}

Figure \ref{fig:transient_cf} shows the friction factor predictions from the Brunone model alongside RANS simulation results for the test cases outlined in Table \ref{tab:ramp_up_down}. The following key observations can be made:
\begin{itemize}
    \item The friction factor evolution differs significantly between ramp-up and ramp-down flows, highlighting the asymmetry in transient shear stress behaviour.
    \item For moderate transients ($t=1s$ and $t=0.1s$), the quasi-steady friction model remains a reasonable approximation, closely aligning with the RANS predictions.
    \item Noticeable deviations from quasi-steady behaviour appear only in cases of stronger transients ($t=0.01s$), where transient effects become dominant regardless of whether the flow is accelerating or decelerating.
\end{itemize}

In strong ramp-up flows, the friction factor $C_f$ exhibits a significant overshoot compared to its quasi-steady counterpart during the initial phase of acceleration. The Brunone model successfully captures such behaviour in all three reference temperatures investigated, which is encouraging. However, in the later phase of the transient, $C_f$ undershoots the quasi-steady values due to the delayed response of turbulence, a phenomenon that the Brunone model does not account for, as previously mentioned. In strong ramp-down flows, however, the Brunone model predicts a trend opposite to that observed in RANS simulations, indicating a limitation in its applicability to ramp-down transients.  
\smallskip

In summary, accurately predicting flows with strong transient effects using SubChCFD remains a challenge, largely due to the lack of sophisticated transient friction models. To the best of the authors' knowledge, no existing models can fully capture these complex effects. One-dimensional transient friction models, such as the Brunone model, can partially capture the dynamic response of wall shear stress to flow transients, but their applicability is limited and not universally reliable across different flow conditions. Given these limitations, the transient friction models should be used for specific flow scenarios, with careful consideration of the trade-off between potential accuracy improvement and uncertainties introduced.
\smallskip

\subsection{Variable property correction}
Nuclear reactors normally involve high-temperature variations and high heat fluxes, which may give rise to substantial changes in coolant physical properties, such as density, viscosity, thermal conductivity, and specific heat capacity, impacting significantly the thermal-hydraulic behaviour of the system. Traditional friction and heat transfer correlations, derived based on constant fluid property assumptions, must be modified to account for these variations. This is particularly critical for supercritical and near-critical fluids, where sharp property variations near pseudo-critical temperatures significantly alter turbulence structures and heat transfer mechanisms.
In one of our previous studies \cite{zhang2023new}, a heat transfer correlation incorporating variable property effects \cite{he2008assessment} has been implemented in SubChCFD to enhance the prediction accuracy of the tool for heat transfer of supercritical fluids. The correlation is formulated by modifying the classic Dittus-Boelter Nusselt correlation using correction coefficients derived from the wall-to-bulk property ratios ( e.g. $(\rho_w/\rho_b)^m$, with $\rho_w$ and $\rho_b$ being the wall and bulk densities, respectively):

\begin{equation} \label{eq:Nu_sc}
    Nu = 0.0183 \, Re_b^{0.82} Pr_b^{0.5} \left( \frac{\rho_w}{\rho_b} \right)^{0.3} \left( \frac{\bar{C_p}}{C_{p_b}} \right)^n
\end{equation}

\noindent
where,

\begin{equation} \label{eq:cp}
    \bar{C_p} = \frac{1}{T_w - T_b} \int_{T_b}^{T_w} C_p \, dT = \frac{h_w - h_b}{T_w - T_b}
\end{equation}

\noindent
The exponent $n$ is determined based on different temperature conditions:

\begin{equation} \label{eq:n}
n =
\begin{cases}
    0.4, & \text{for } T_b < T_w \leq T_{pc} \text{ or } 1.2 T_{pc} < T_b \leq T_w \\
    0.4 + 0.2 \left( \frac{T_w}{T_{pc}} - 1 \right), & \text{for } T_b \leq T_{pc} < T_w \\
    \left[ 0.4 + 0.2 \left( \frac{T_w}{T_{pc}} - 1 \right) \right] \left[1 - 5 \left( \frac{T_b}{T_{pc}} - 1 \right) \right], & \text{for } T_{pc} < T_b \leq 1.2 T_{pc} \text{ and } T_b < T_w
\end{cases}
\end{equation}

\noindent
where $T_w$ is the wall temperature, $T_b$ the bulk temperature, and $T_{pc}$ the pseudo-critical temperature of the working fluid at a certain pressure above the critical pressure.
\smallskip

\begin{figure}[ht]
    \centering
    \begin{subfigure}[t]{0.45\textwidth}
        \centering\includegraphics[width=1\linewidth]{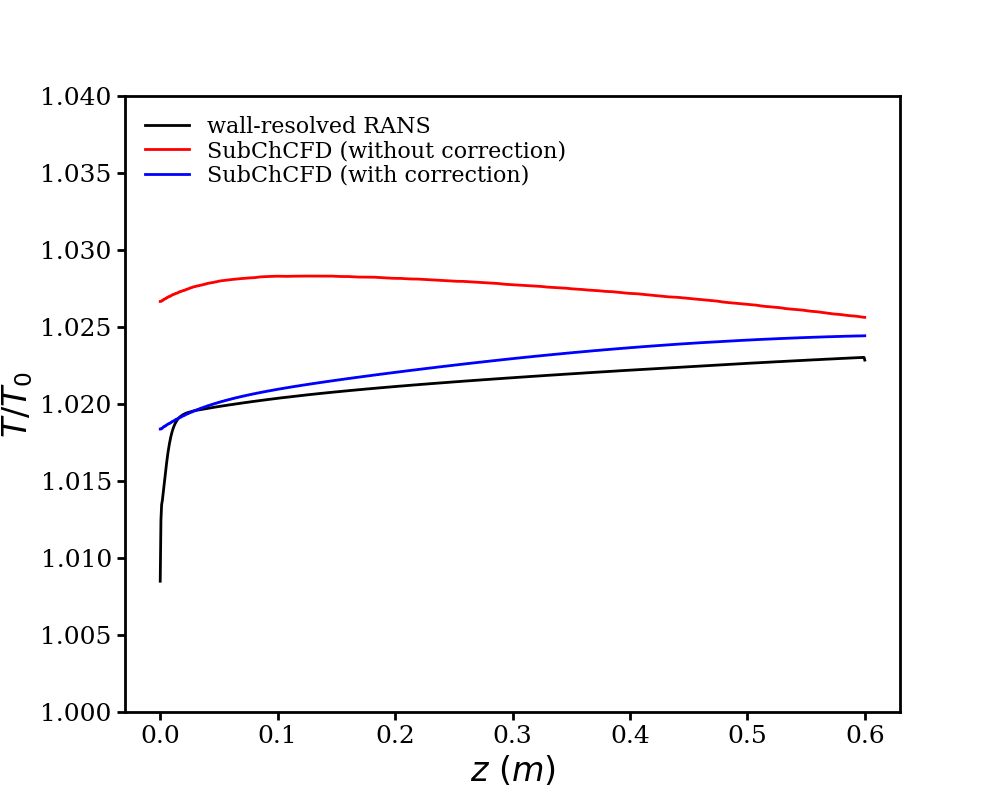}
        \caption{centre channel}
    \end{subfigure}
    \begin{subfigure}[t]{0.45\textwidth}
        \centering\includegraphics[width=1\linewidth]{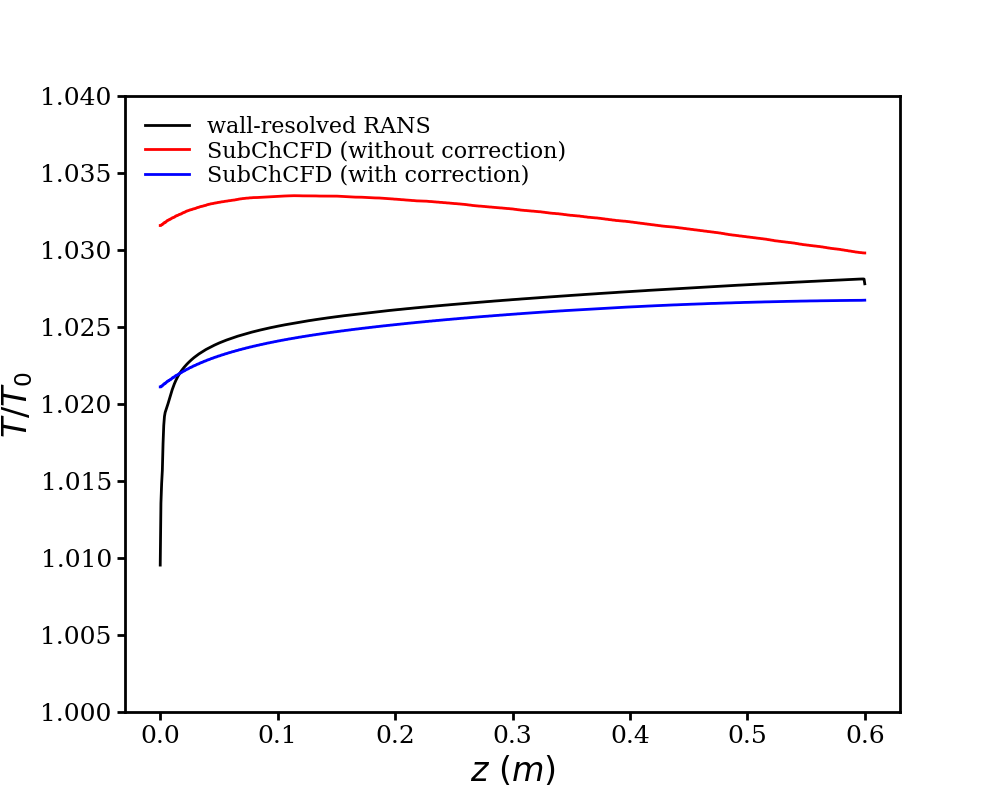}
        \caption{edge channel}
    \end{subfigure}
    \caption{Comparison of SubChCFD-predicted rod surface temperatures in a 2×2 rod bundle with supercritical water. The corrected model (blue) shows closer agreement with reference RANS data (black) than the uncorrected model (red).}
    \label{fig:supercritical}
\end{figure}

Equipped with the heat transfer correlation detailed by Equations \ref{eq:Nu_sc} to \ref{eq:n}, SubChCFD was applied to simulate a 2$\times$2 rod bundle with supercritical pressure water as the working fluid. Figure \ref{fig:supercritical} shows the axial development of the rod surface temperature. It is clearly indicated that significant improvement was achieved in the rod surface temperature predictions when the variable property effects were taken into account in the heat transfer model used. The corrected model (blue) shows much closer agreement with reference RANS data (black) than the uncorrected model (red), which severely over-predicted the rod surface temperature.

In the current project, considering the fact that HTGRs operate at very high temperatures, and large temperature gradients can develop due to localised poor heat transfer in natural circulation during a LOFA event, variable property effects need to be suitability taken into account. In this report, we demonstrate the implementation of an additional correlation, the Petukhov correlation \cite{polyakov1991heat}. This correlation has been extensively validated against experimental data and covers a broad range of heat transfer conditions. More importantly, it is designed for sub-critical conditions, more relevant to HTGR-related applications. Additionally, the methodology detailed here will shed light on the endeavours of potential SubChCFD users, who aim to implement their own correlations following a similar approach.
\smallskip

The Petukhov correlation corrects both the skin friction factor and Nusselt number simultaneously. The implemented version in SubChCFD involves a few calculation steps. In \textbf{Step 1}: the subchannel bulk flow parameters ($U_b$, $\rho_b$, $\mu_b$, $\lambda_b$ and $C_{p,b}$), obtained from initialisation (for the first time step) or the previous iteration/time step, are used to compute subchannel Reynolds and Prandtl numbers, with which an initial estimate of the friction factor can be obtained as:

\begin{equation} \label{eq:Cf_uncorrected}
    C_{f0} = \frac{1}{\left[ 1.82 \log_{10} \left( \frac{Re_b}{8} \right) \right]^2}
\end{equation}

\noindent
In the case of taking into account transient effects, the Brunone model described in Section \ref{sec:development:subsection1} can be used as a replacement for Equation \ref{eq:Cf_uncorrected}. In \textbf{Step 2}, bulk-to-wall ratios of density and viscosity are used as correction coefficients to incorporate property effects:

\begin{equation} \label{eq:Cf_corrected}
    C_f = C_{f0} \left( \frac{\rho_w}{\rho_b} \right)^m \left( \frac{\mu_w}{\mu_b} \right)^n
\end{equation}

\noindent
where $\rho_w$ and $\mu_w$ are density and viscosity based on the wall temperatures (also coming from initialisation or previous iteration). The exponents $m$ and $n$ are selected based on Petukhov’s work, with recommended values $m=0.4$ and $n=1.0$. Alternative values, such as $m=0.33$ and $n=0.2$, have also been suggested based on large datasets of different fluid conditions by other researchers \cite{polyakov1991heat}. Users can select their preferred values accordingly. In \textbf{Step 3}, the Nusselt number is recalculated with the corrected friction factor $C_f$ as, 

\begin{equation} \label{eq:Nu_corrected}
    Nu = \frac{Re_b Pr_b \left( \frac{C_f}{8} \right)}
         {1 + \frac{900}{Re_b} + 12.7 \left( \frac{C_f}{8} \right)^{0.5} (Pr_b^{2/3} - 1)}
\end{equation}

\begin{figure}[H]
    \centering\includegraphics[width=0.8\linewidth]{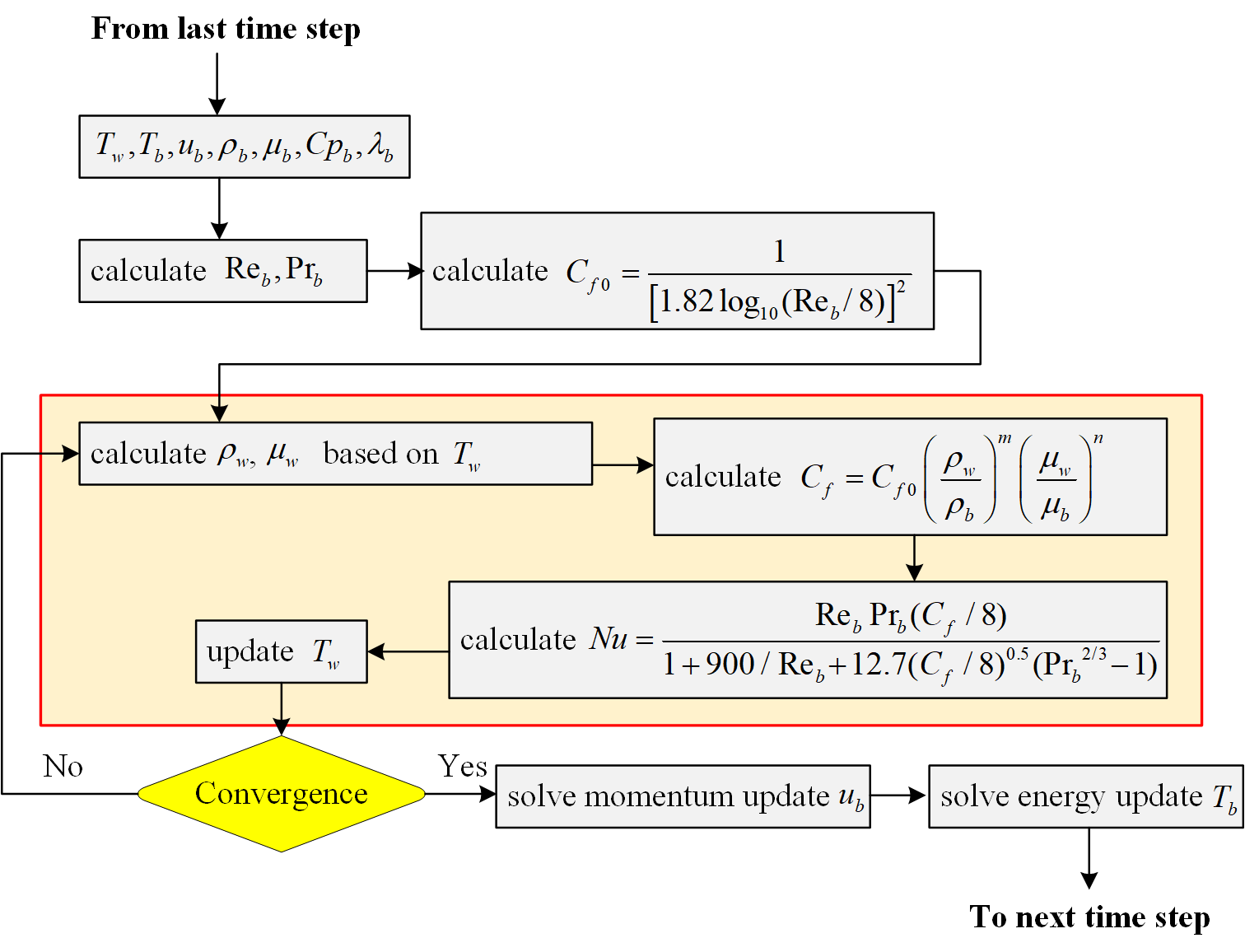}
    \caption{Workflow for implementation of the Petukhov correlations into SubChCFD.}
    \label{fig:petukhov}
\end{figure}

\noindent
Likewise, further corrections can be applied to the Nusselt number using the bulk-to-wall property ratios, which are not detailed here. In \textbf{Step 4}, the wall temperature is updated using Newton’s cooling law for each local subchannels, based on the Nusselt number computed. This updated wall temperature is then feedback to \textbf{Step 2}, to update the correction factors (the wall-to-bulk property ratios) in Equation \ref{eq:Cf_corrected}. \textbf{Step 2} to \textbf{Step 4} are repeated until a convergent solution of wall temperature is obtained, when the difference between adjacent iterations being lower than a certain tolerance, $||T_w^{n+1}-T_w^n||<\epsilon$. Then, the simulation leaves the current time step and proceeds to the next time step. Figure \ref{fig:petukhov} outlines the workflow for integrating Petukhov’s correlations into SubChCFD, illustrating the calculation steps described above.

\begin{figure}[H]
    \centering\includegraphics[width=0.8\linewidth]{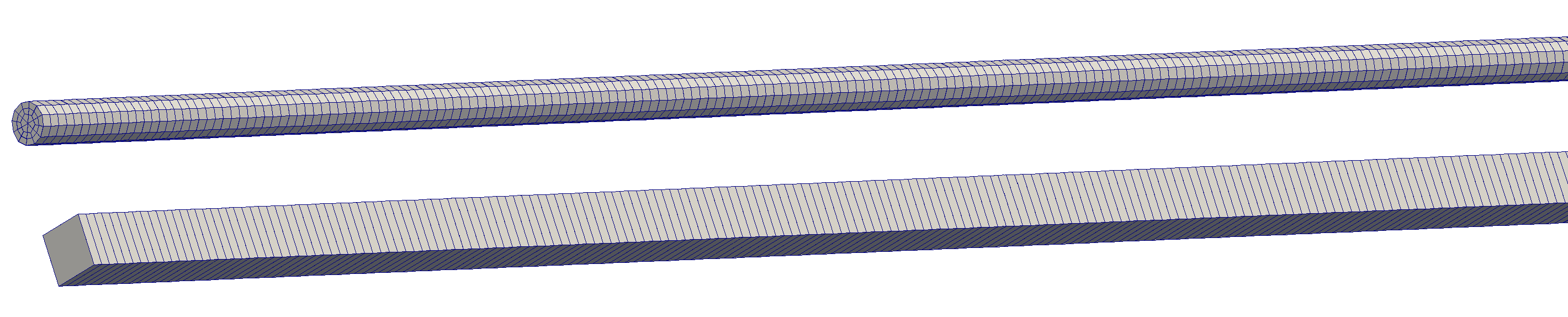}
    \caption{Coarse-grid computing and filtering meshes used for the SubChCFD modelling of a pipe flow.}
    \label{fig:coarse_pipe}
\end{figure}

For a rapid verification of the implemented model, SubChCFD has been adapted with the capability for modelling pipe flows. The selected pipe is 7.93 m long and 15.88 mm in diameter, mimicking the geometry of a single coolant channel in a prismatic HTGR core. Again, helium is used as the working fluid, the same as the coolant used in HTGRs. The pipe walls are uniformly heated with a heat flux of $1.46\times10^5$ W/m², while the inlet velocity of helium is set at 28 m/s and inlet temperature of 490 $^\circ C$, estimated based on HTGR operational conditions. Figure \ref{fig:coarse_pipe} shows the computational meshes used in the SubChCFD model. A coarse-grid computing mesh is created, consisting of 30 cells in the cross-section and 1,895 divisions along the axial direction. On top of that, a filtering mesh, maintaining the same axial resolution as the computing mesh, is used to produce cross-sectional average flow quantities. To provide a reference for validation of the SubChCFD predictions, a detailed CFD simulation using the $k-\omega$ SST turbulence model is conducted for the same geometry and thermal-hydraulic conditions.
\smallskip

\begin{figure}[H]
    \centering\includegraphics[width=0.8\linewidth]{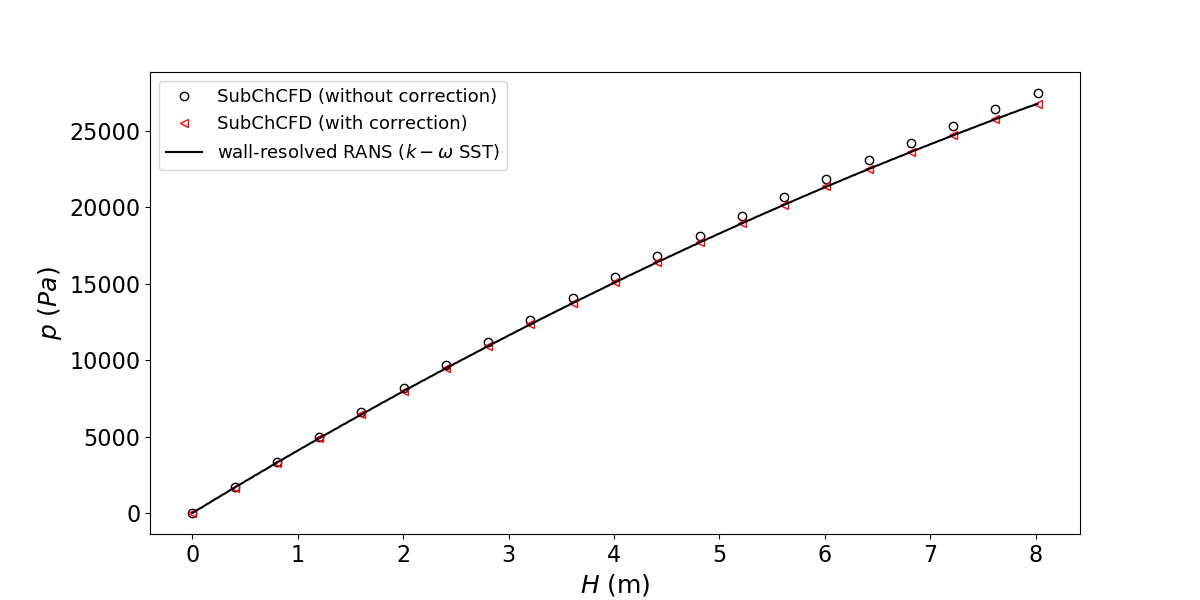}
    \caption{Comparison of the predicted pressure drops in a pipe flow.}
    \label{fig:pressure_drop_pipe}
\end{figure}

Figure \ref{fig:pressure_drop_pipe} shows the pressure drop results. The SubChCFD predictions without variable property corrections slightly overestimate the overall pressure drop compared to the reference CFD results. This suggests that despite the high heating power, the variable property effect is not as dominant as initially expected. However, after implementing variable property corrections, a noticeable improvement in SubChCFD predictions is observed. The axial pressure distribution closely matches the reference CFD data, confirming that the Petukhov correlation accurately predicts wall friction under property variation effects. For heat transfer, the effects of property variation in this particular case are minimal, so the results are not presented here.

\subsection{Buoyancy correction}
Following the onset of a LOFA in an HTGR core, the primary heat transfer mechanism transitions from forced convection to natural circulation due to the rapid reduction in forced flow after the pump trips, causing the flow to reverse direction. In the absence of forced flow, helium circulation within the core is mainly driven by the buoyancy force arising from density differences across the coolant channels. To accurately simulate this process, it is essential for the heat transfer models in SubChCFD to take into account the buoyancy effect.
\smallskip

Buoyancy-induced alteration to turbulence and heat transfer have been extensively studied over the past few decades. General consensus has been reached on  understandings of the fundamental phenomena. In upward (buoyancy-aided) flows, moderate buoyancy effects typically impair heat transfer, while strong buoyancy effects enhances heat transfer, known as heat transfer recovery. In downward (buoyancy-opposed) flows, heat transfer is consistently enhanced as buoyancy effects increase \cite{marensi2021}. To quantify the influence of buoyancy on heat transfer, Jackson and Hall \cite{Jackson1979} introduced a non-dimensional group, the buoyancy parameter, to  characterise the extent of buoyancy influence in mixed and natural convection flows. The non-dimensional group is defined as:

\begin{equation} \label{eq:Bo}
    Bo^* = \frac{Gr^*}{Re^{3.425} Pr^{0.8}}
\end{equation}  

\noindent
where $Gr^*$ is a modified Grashof number, $Re$ is the Reynolds number, and $Pr$ is the Prandtl number. By relating the shear stress in the near-wall region to heat transfer, a semi-empirical model was developed to relate the mixed convection Nusselt number to the buoyancy parameter. The model is formulated as:

\begin{equation} \label{eq:NubyNuf}
    \frac{Nu}{Nu_f} = \left( 1 \mp 2.5 \times 10^5 Bo^* \left( \frac{Nu}{Nu_f} \right)^{-2} \right)^{0.46}
\end{equation}

\noindent
where, the minus and plus signs correspond to the buoyancy-aided and buoyancy-opposed flows, respectively. $Nu$ represents the Nusselt number of the buoyancy-influenced flow, while $Nu_f$ is the Nusselt number in the absence of buoyancy effects. In SubChCFD, upon obtaining $Nu_f$ (through Equations \ref{eq:Nu_sc} or \ref{eq:Nu_corrected}), the Nusselt number ratio $Nu/Nu_f$ in Equation \ref{eq:NubyNuf} can be used effectively as a correction coefficient to account for the buoyancy effects.
\smallskip

To optimise the applicability of this model for HTGR-relevant scenarios, the model parameter in Equation \ref{eq:NubyNuf} was  calibrated slightly. The calibration was based upon RANS simulations due to the lack of experimental and/or DNS data. The Launder-Sharma $k-\epsilon$ turbulence model was used, as it was found to perform the best among various RANS turbulence modes in capturing heat transfer deterioration and enhancement in buoyancy-influenced flows \cite{kim2008}. To isolate the buoyancy effect, the Boussinesq approximation was used, neglecting all property variations except density in the gravity term. Consequently, the model was simplified to a 2-D representation using an axisymmetric coordinate system, with periodicity imposed in the flow direction. The investigated cases were selected based on expected thermal-hydraulic conditions of the coolant channels relevant to a LOFA event, covering both upward and downward flow scenarios. The key parameters of these cases are summarised in Table \ref{tab:conditions}.

\floatstyle{plaintop}
\restylefloat{table}
\begin{table}[h]
    \centering
    \renewcommand{\arraystretch}{1.0}
    \setlength{\tabcolsep}{8pt} 
    \begin{tabular}{c | c | c | c}
        \hline\hline
        \textbf{$U$ (m/s)} & \textbf{$\dot{Q}$ (kW/m\textsuperscript{2})} & \textbf{$Re$} & \textbf{$Bo^*(\times10^{-6})$} \\
        \hline
        2.89 & 29.1   & 5134 & 0.26 \\
        2.89 & 72.8   & 5134 & 0.65 \\
        2.89 & 145.5  & 5134 & 1.30 \\
        2.89 & 218.3  & 5134 & 1.95 \\
        2.32 & 145.5  & 4108 & 2.79 \\
        1.74 & 218.3  & 3080 & 11.20 \\
        \hline\hline
    \end{tabular}
    \caption{Flow conditions.}
    \label{tab:conditions}
\end{table}

\begin{figure}[H]
    \centering
    \begin{subfigure}[t]{0.45\textwidth}
        \centering\includegraphics[width=1\linewidth]{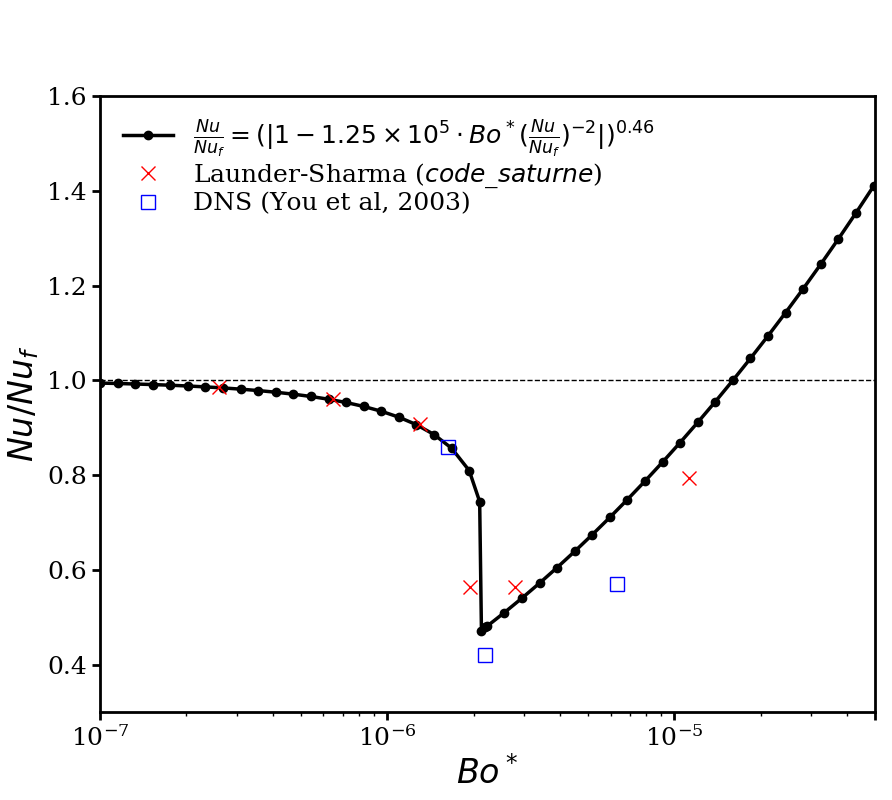}
        \caption{upward flow}
    \end{subfigure}
    \begin{subfigure}[t]{0.45\textwidth}
        \centering\includegraphics[width=1\linewidth]{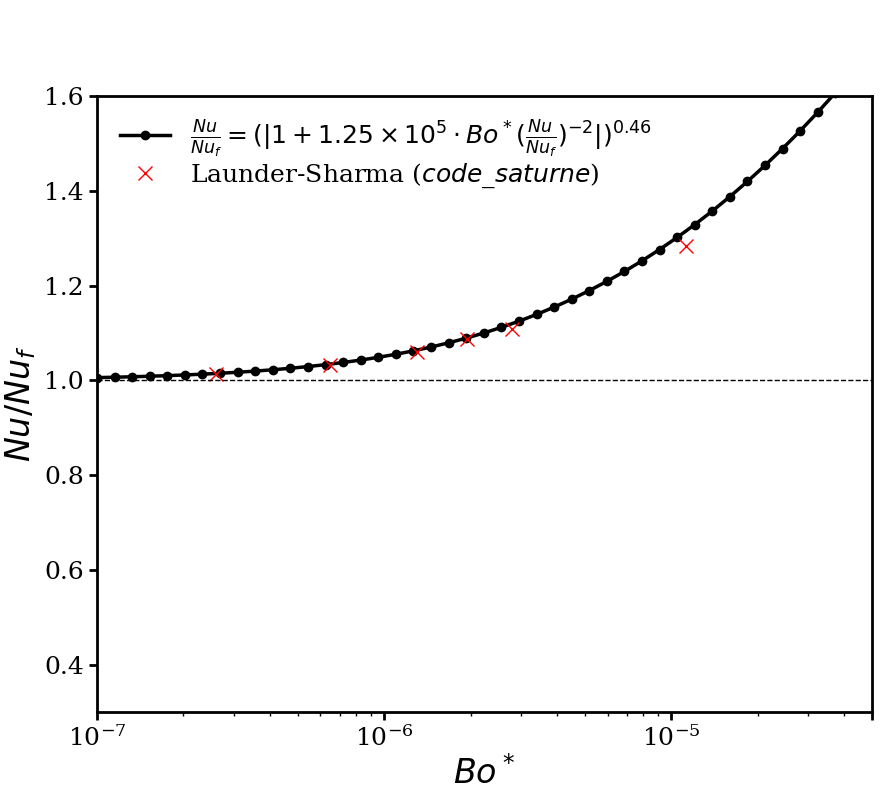}
        \caption{downward flow}
    \end{subfigure}
    \caption{Calibrated $Nu/Nu_f-Bo^*$ correlation for buoyancy correction.}
    \label{fig:Nu_table}
\end{figure}

The calibration is prioritised for moderate buoyancy-influenced flows, considering the fact that it is unlikely the buoyancy effects can be strong enough to induce heat transfer recovery in the context of LOFA conditions in HTGRs, given that the decay heat power is expected to decrease immediately after the onset of LOFA, remaining below 10\% of nominal conditions. Overall, only minimal calibration was required, with the constant $2.5\times10^5$ in Equation \ref{eq:NubyNuf} adjusted to $1.25\times10^5$, which was found to be in best agreement with the RANS simulation results. Figure \ref{fig:Nu_table} shows the calibrated $Nu/Nu_f-Bo^*$ relationship. For upward flows, the calibrated correlation closely aligns with the RANS results as well as some DNS data from literature for the left branch ($Bo^*\leq2\times10^{-6}$), which corresponds to heat transfer deterioration due to moderate buoyancy influence. However, some deviation is allowed for the right branch ($Bo^*>2\times10^{-6}$) as a compromise, which corresponds to the heat transfer recovery regime under strong buoyancy. For downward flows, the calibrated model aligns consistently with the reference data.
\smallskip

In SubChCFD, $Bo^*$ is calculated locally for each subchannel at each time step, and $Nu/Nu_f$ is computed accordingly to update the heat transfer coefficient. Instead of solving non-linear equations to get $Nu/Nu_f$, look-up tables are created beforehand and incorporated in SubChCFD to reduce computational effort.

\section{Modelling of LOFA in an HTGR using a 1/12th core model}
\label{sec:modelling}
\subsection{The SubChCFD model}
\subsubsection{Computational domain}
Unlike steady-state conditions, where symmetry assumptions can be applied to small-scale models such as individual fuel blocks due to the relatively uniform thermal environment, LOFA transients introduce large-scale non-uniform temperature distributions across the core. These temperature variations result in complex, asymmetric flow patterns that significantly impact the transport of decay heat and thus the thermal behaviour of the whole reactor system. Since natural convection becomes the dominant heat transfer mechanism after forced flow ceases, accurately modeling the buoyant flow paths in the upper and lower plena and throughout the core is essential. Smaller computational domains, such as a 1/12th sector model, enforce excessive symmetry constraints, preventing the simulation from capturing the natural circulation loops realistically. Such models restrict the natural development of buoyancy-driven flow structures, leading to errors in temperature distribution and flow behaviour predictions. On the other hand, while a full-core model would be the most accurate approach, it is still computationally expensive even a coarse-grid is used. Therefore, the 1/12th core model is adopted as an optimal compromise, leveraging the symmetric features of the core while retaining enough spatial coverage to capture asymmetries in the transient thermal-hydraulic response.
\smallskip

\begin{figure}[H]
    \centering\includegraphics[width=1.0\linewidth]{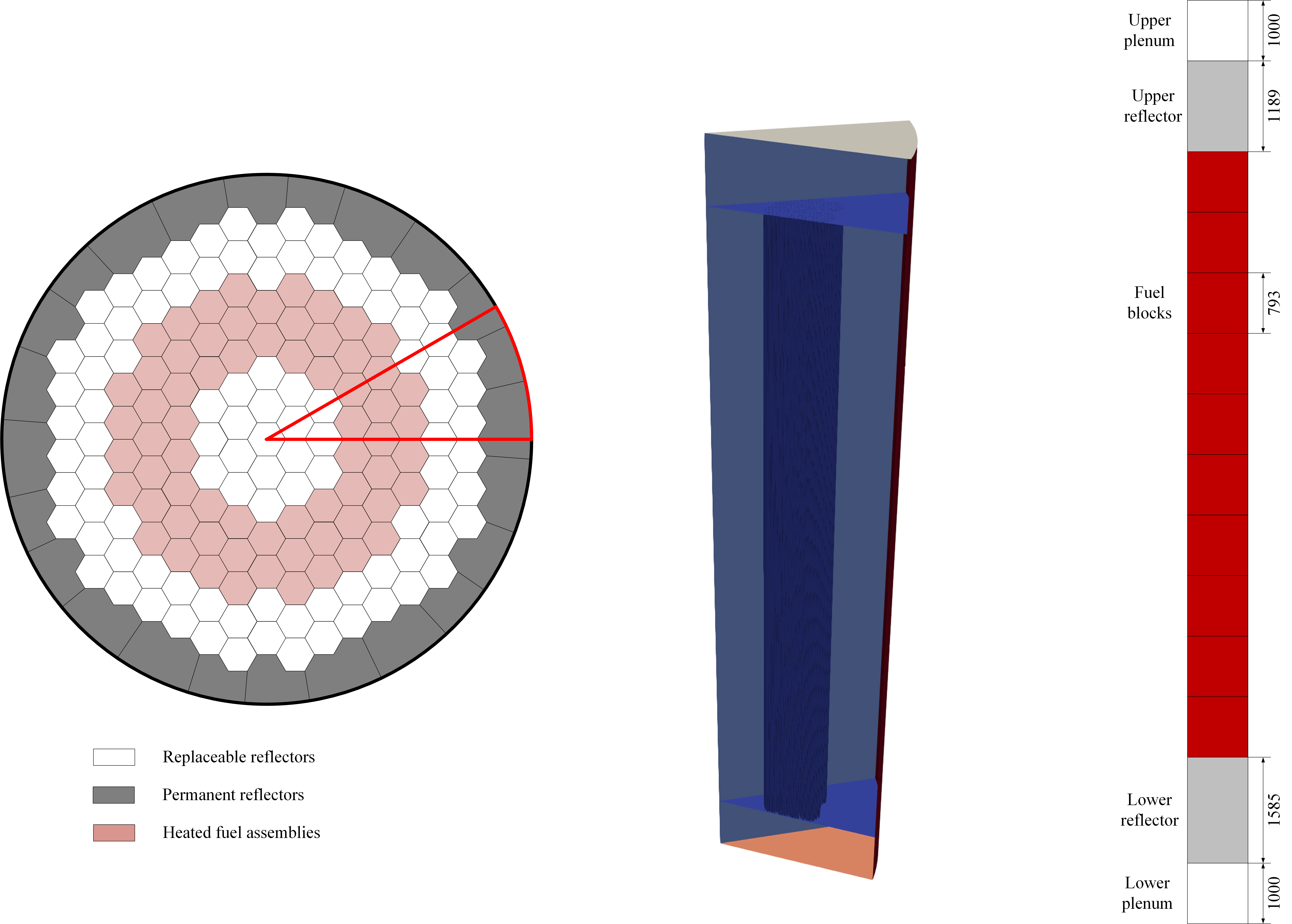}
    \caption{Computational domain. Left: top-down cross-sectional view of the core with the 1/12th core model highlighted, Middle: 3-D representation of the 1/12th core model, Right: height profile of the computational domain.}
    \label{fig:full_core_sketch}
\end{figure}

Figure \ref{fig:full_core_sketch} shows a schematic representation of an typical HTGR core with the 1/12th core model computational domain highlighted. The left sub-figure is a top-down cross-sectional view of the reactor core, where different zones are colour differently to distinguish the key regions. The heated fuel assemblies are shown in red, located in an annular region within the core, surrounded by the inner and outer replaceable reflectors, which are coloured in white. The permanent reflectors, coloured in dark gray, form an outer layer that serves as neutron shielding and provides structural integrity to the core. In the 1/12th core model, there are totally 5.5 heated fuel assemblies simulated. For simplicity, these are all regular assemblies (without control rod holes), and the bypass gaps in between the fuel assemblies are not considered.
\smallskip

The middle image of Figure \ref{fig:full_core_sketch} presents a 3-D representation of the 1/12th core model, showing the functional components of the reactor core. The upper and lower reflectors are considered, which are connected to the heated fuel assemblies at the two ends. These reflectors are unheated but fabricated with coolant channels to ensure continuous flow paths for the coolant through the core. The lower and upper plena are also considered with the dome-shaped pressure boundaries replaced with flat surfaces to simplify the mesh generation. Additionally, all internal components in the upper and lower plena, such as control rod assemblies, support structures, are ignored. Figure \ref{fig:full_core_sketch} (the right image) also shows detailed dimensional information of these components, which are further summarised in Table \ref{tab:computational_domain}. All dimensions, except for the plena, are based on the General Atomics modular HTGR design \cite{NEA2017}. Tung et al. \cite{Tung2016} demonstrated that using simplified plena with a length of 1 m is sufficient for LOFA simulations, as numerical results become insensitive to further increases in plenum length. Consequently, both the upper and lower plena are set to 1 m in length to optimise computational efficiency while maintaining accuracy.
\smallskip

\floatstyle{plaintop}
\begin{table}[H]
    \centering
    \renewcommand{\arraystretch}{1.0}
    \begin{tabular}{l c l}
    \hline\hline
    \textbf{Component} & \textbf{Length (m)} & \textbf{Description}   \\
    \hline
    Upper plenum    & 1.0   & Region where helium coolant enters      \\ 
                    &       & before passing through the core         \\
    Upper reflector & 1.189 & Reflects neutrons back into the core,   \\
                    &       & improving fuel utilisation              \\
    Heated section  & 7.93  & Main heat-generating section            \\
                    &       & consisting of 10 heated fuel blocks     \\
    Lower reflector & 1.585 & Reflects neutrons back upward and       \\ 
                    &       & provides structural support             \\
    Lower plenum    & 1.0   & Region where helium coolant exits       \\ 
                    &       & after extracting heat from the core     \\
    \hline\hline
    \end{tabular}
    \caption{Description of the computational domain with key dimensions of the components}
    \label{tab:computational_domain}
\end{table}

\subsubsection{Thermal physical properties}
Due to the significant temperature variations within the reactor core during a LOFA, the thermal physical properties of both the solid and fluid must be treated as temperature-dependent to ensure accurate heat transfer modeling. In this study, the helium gas properties are tabulated at 1 $^\circ$C intervals under a constant pressure of 7 MPa (operating pressure of an HTGR) using data from the NIST database REFPROP v9 \cite{lemmon2010nist}. Additionally, the thermal physical properties of graphite and fuel compacts are derived from Tak et al. \cite{Tak2008} and Johnson et al. \cite{Johnson2009} and re-fitted using fourth-order polynomials \cite{Liu2023} for ease of implementation in the solver.

\subsubsection{Initial condition of LOFA}
\begin{figure}[H]
    \centering\includegraphics[width=0.45\linewidth]{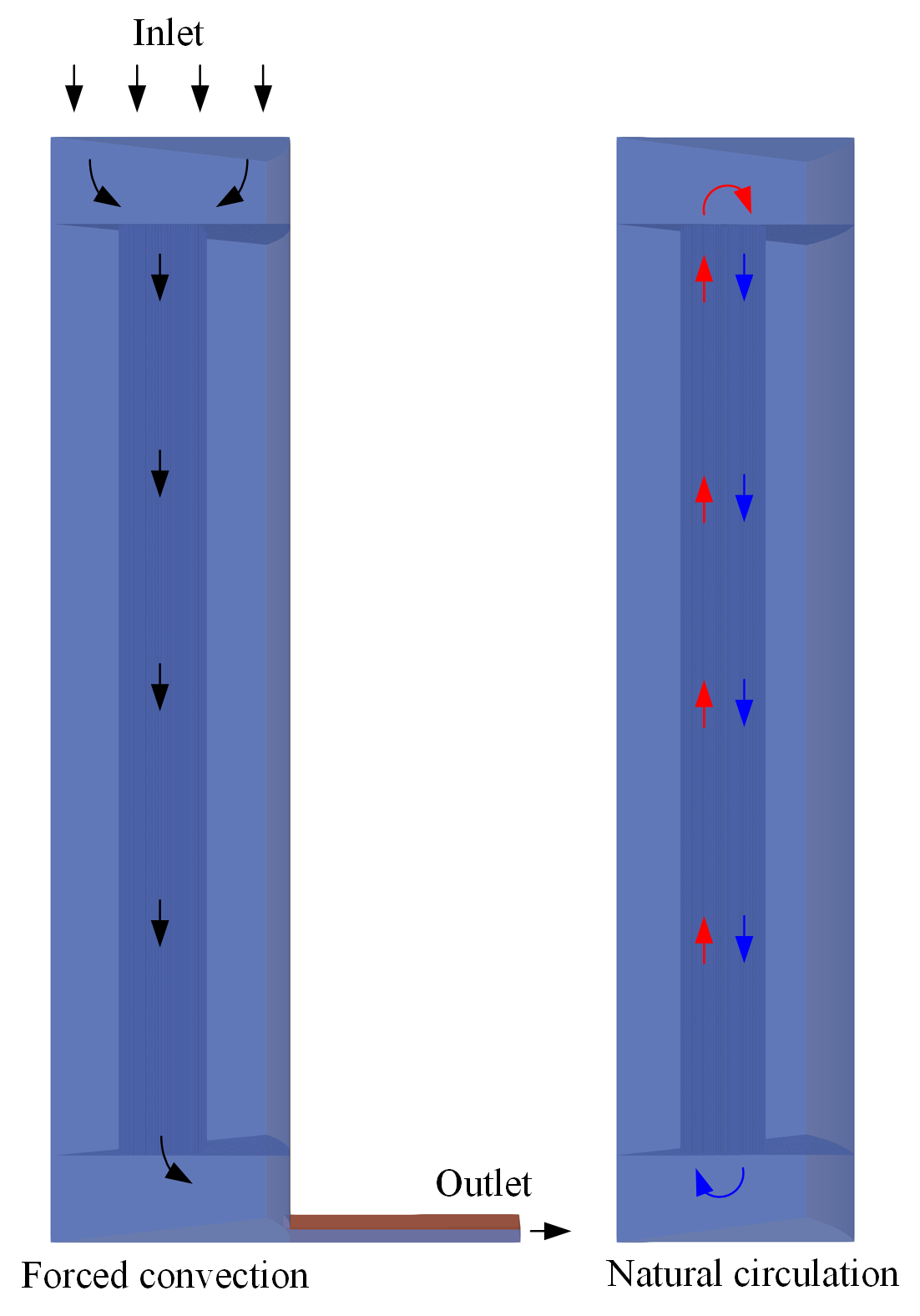}
    \caption{Initialisation of the LOFA simulation and transition from forced convection to natural circulation in an HTGR core. Left: computational domain with inlet and outlet for steady-state operational condition, Right: computational domain for the LOFA transient simulation.}
    \label{fig:domain_comparison}
\end{figure}

The LOFA transient starts from an initial state with well established velocity and temperature fields based on operational conditions of an HTGR. To obtain this initial condition, a steady-state SubChCFD simulation was conducted using the same computational domain modified slightly to have an inlet and an outlet, as shown in Figure \ref{fig:domain_comparison} (left image). The inlet is created by directly making an opening at the top of the upper plenum, and a mass flow inlet condition is used. While, the outlet is through a horizontal duct added to one side of the lower plenum to mimic the coolant passage connecting the core and the steam generator, with a constant pressure imposed.
\smallskip

The LOFA event is triggered by changing the flow boundary conditions to reflect the sudden loss of coolant circulation. At the onset of the transient simulation, the mass flow inlet is replaced with a no-slip wall and the outlet duct is removed to seal the system (see Figure \ref{fig:domain_comparison} (right image)), effectively halting forced flow and allowing natural circulation to govern helium movement within the core. Such modification results in a pressure drop across the core that quickly approaches zero, forcing the transition from forced convection to buoyancy-driven flow. Additionally, the heat generation rate is reduced to $\sim$ 10\% of its nominal value to simulate the reactor’s decay heat power following a reactor shutdown. Table \ref{tab:conditions1} compares detailed parameters and settings used in the steady-state and transient models. 

\floatstyle{plaintop}
\restylefloat{table}
\begin{table}[H] 
    \centering
    \renewcommand{\arraystretch}{1.0}
    \begin{tabular}{l l l}
    \hline\hline
    & \textbf{Steady state} & \textbf{Transient} \\
    \hline
    Inlet              & Mass flow inlet  & non-slip wall \\ 
    - Flow rate        & 14.35 kg/s  &   \\
    - Temperature      & 490 $^\circ$C  &   \\
    Outlet             & Fixed pressure at 7 MPa  & non-slip wall  \\
    Core outer surface & Fixed temperature at 490 $^\circ$C  & Fixed temperature at 490 $^\circ$C  \\
    Power density      & 3.11$\times10^7$ W/m\textsuperscript{3} & $\leq$ 3.11$\times10^6$ W/m\textsuperscript{3}  \\ 
    \hline\hline
    \end{tabular}
    \caption{Model parameters and settings for the steady-state and the transient simulations.}
    \label{tab:conditions1}
\end{table}

\subsubsection{Cases simulated}
\label{sec:cases}
During a LOFA event, since helium motion is no longer actively driven, the flow rates decrease substantially, and the heat removal capacity is significantly reduced. This can lead to the development of localised hot spots within the core, posing challenges to the safety and structural integrity of the reactor. The rate at which decay heat diminishes over time directly impacts the temperature evolution within the core. If decay heat remains high for an extended period, the core can experience significant temperature buildup, potentially leading to hot spots and structural material degradation. Conversely, a rapid decay heat reduction allows for a more effective passive cooling response, minimising excessive temperature rises.
\smallskip

\begin{figure}[H]
    \centering\includegraphics[width=0.6\linewidth]{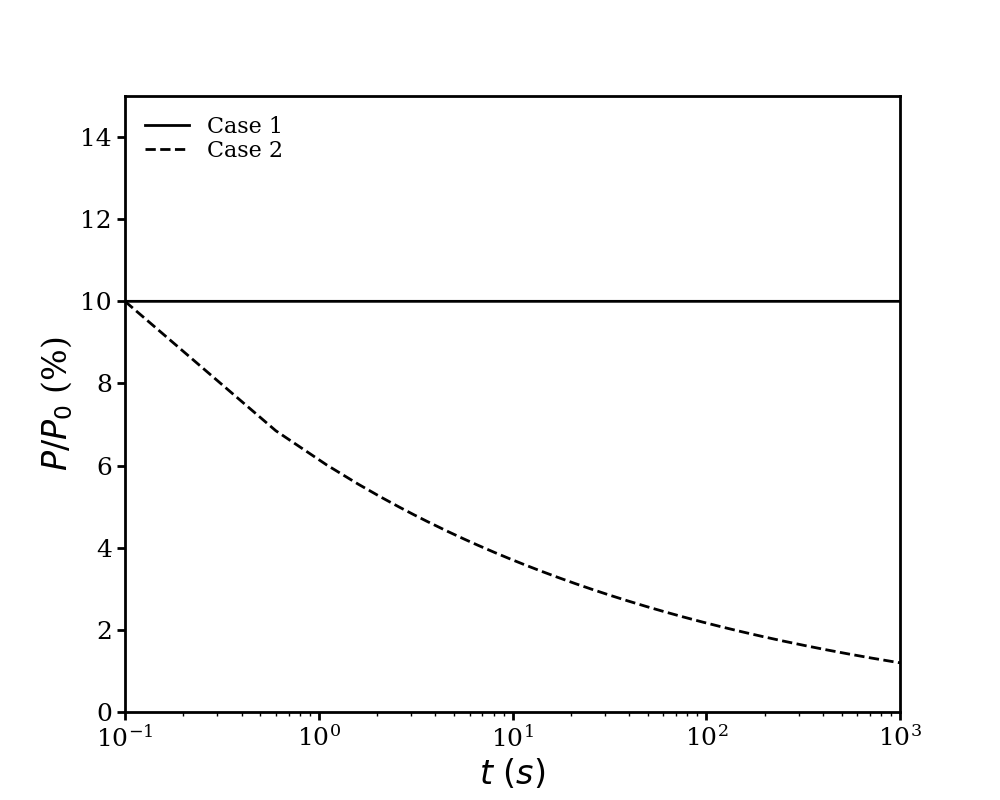}
    \caption{Decay heat power histories of the investigated cases.}
    \label{fig:power_history}
\end{figure}

In this study, we considered two distinct decay heat power histories for our LOFA simulations, as illustrated in Figure \ref{fig:power_history}. Case 1 represents a conservative scenario where decay heat remains at a constant 10\% of nominal power. While, case 2 represents a more realistic scenario consistent with fission product decay, where power decreases over time following the below correlation \cite{Todreas1990}:

\begin{equation} \label{eq:PbyP0}
    P/P_0 = 0.066\cdot[t^{-0.2}-(t+\tau)^{-0.2}]
\end{equation}

\noindent
where $P$ represents the decay heat power, $P_0$ is the nominal power at steady operation, $t$ is the time after the onset of LOFA, and $\tau$ is a constant relevant to the steady operation time of the reactor prior the occurance of LOFA. 

\subsubsection{Meshing strategy}
\begin{figure}[H]
    \centering\includegraphics[width=0.9\linewidth]{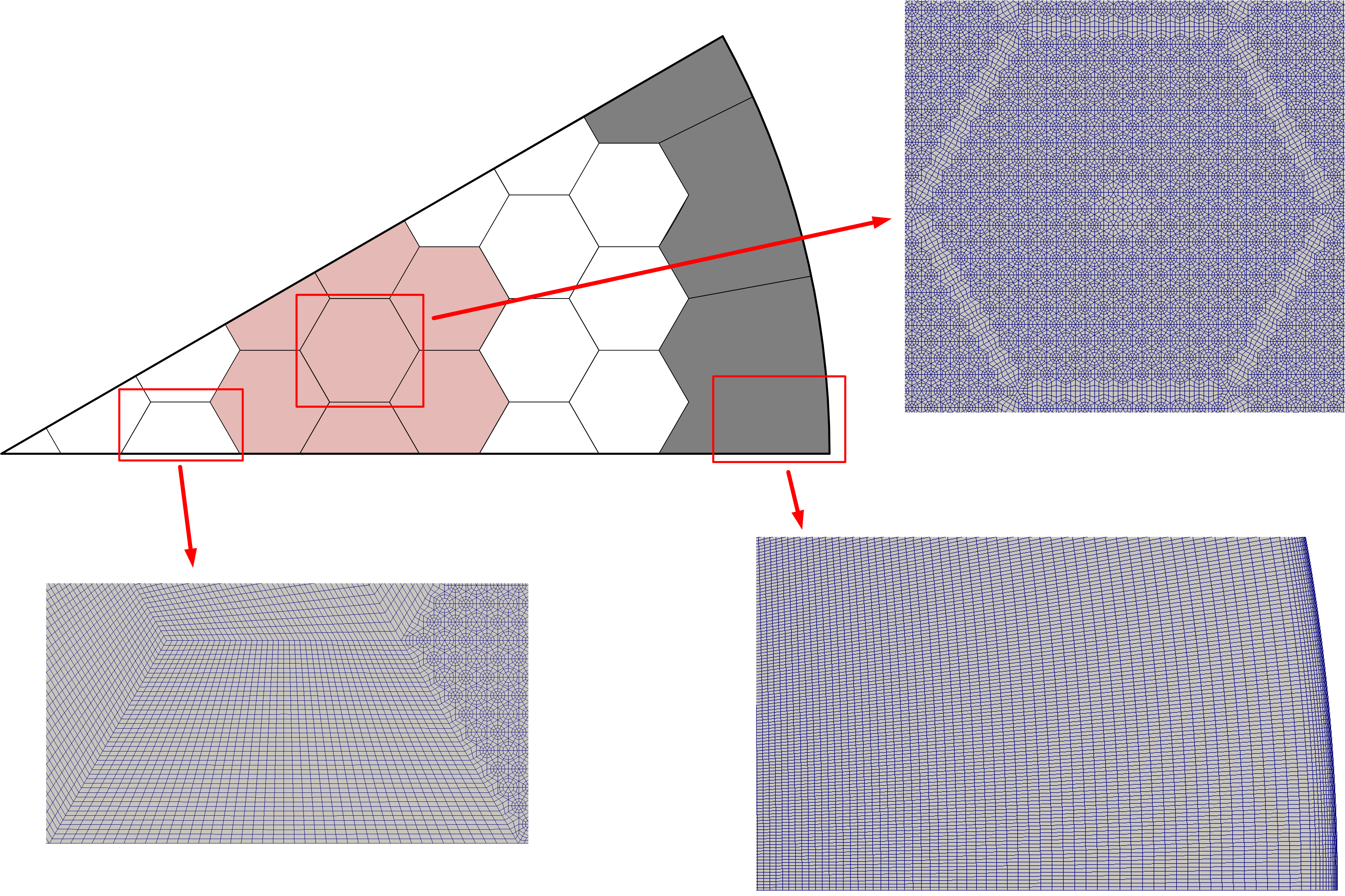}
    \caption{Coarse-grid computing mesh for the 1/12th core model.}
    \label{fig:computing_mesh}
\end{figure}

Figure \ref{fig:computing_mesh} shows a cross-sectional view of the coarse-grid computing mesh for the 1/12th core model. Since the mesh was built primarily through extrusion, the basic mesh topology remains the same across different components. The mesh consists of approximately 63 million computational cells, with around 43 million dedicated to the solid regions. A dedicated internal coupling approach was used to handle the conjugate heat transfer between the fluid and the solid \cite{Liu2023}. Overall, the mesh is slightly finer in the heated fuel blocks and coarser deep within the reflectors where lower temperature gradients are expected.

\begin{figure}[ht]
    \centering\includegraphics[width=0.9\linewidth]{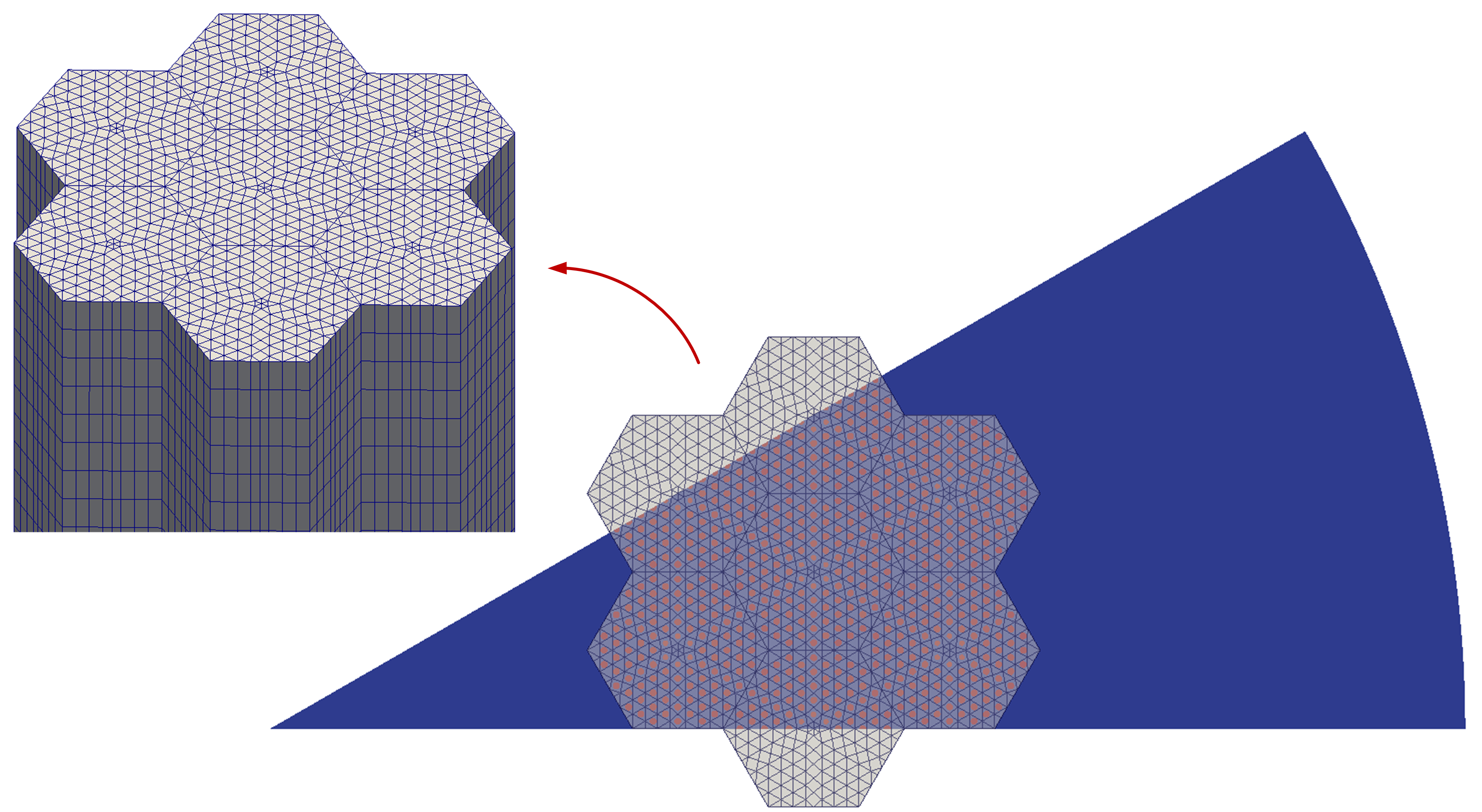}
    \caption{Subchannel mesh for the 1/12th core model.}
    \label{fig:subchannel_mesh}
\end{figure}

Figure \ref{fig:subchannel_mesh} shows the design of the subchannel mesh used in SubChCFD for the 1/12th core model. It can be seen that the subchannel mesh is applied only within the fuel block regions, where coolant channels exist. A key requirement on the subchannel mesh is to ensure an one-to-one correspondence of the subchannel mesh cells with the coolant channels in the computing mesh. Empty cells that do not correspond to any coolant channels are removed through preprocessing. To further improve computational efficiency, the subchannel mesh is designed to be four times coarser in the flow direction compared to the computational mesh, which effectively reduces the total number of subchannel mesh cells and therefore the computational overhead added for computing the average quantities. This is not expected to degrade the accuracy of the simulation, since the flow properties do not vary sharply along the axial direction of the reactor core.

\subsubsection{Numerical simulation}
The numerical simulations were performed on UK's Tier-1 HPC system, ARCHER2, with 32 computing nodes (4,096 CPU cores) used. The objective of the simulations were to predict the transient thermal-hydraulic response of the HTGR core following a LOFA for a physical time at an order of 1000 seconds after the onset of the event. This extended simulation period was essential for understanding the early evolution of natural circulation, heat redistribution, and fuel temperature evolution within the core.
\smallskip

Typically, time step sizes used in the transient simulations were dynamically adjusted to maintain an average CFL number below 0.5. Consequently, each full simulation required approximately 3,000 node hours to research the target physical time. This is significantly lower than that of a convectional RANS approach (see Section \ref{sec:comparisonRANS}), which typically demand substantially longer runtime due to their higher mesh resolution requirements.

\subsection{Results and discussion}
This section presents the SubChCFD simulation results for 
the two cases outlined in Section \ref{sec:cases}, focusing on the velocity and temperature distributions within the reactor core, which highlights the key thermal-hydraulic phenomena governing the evolution of the system under passive cooling conditions. To facilitate examination of these results, a set of probe locations and line plots were defined, enabling detailed analysis of temporal evolution and spatial distribution of velocity and temperature fields within the heated section. As shown in Figure \ref{fig:probes}, probes 1 to 6 are positioned at the centre of coolant channels within three distinct fuel assemblies, each located at different positions within the reactor core. Probes 7 to 9 are situated at the center of fuel compact holes adjacent to the coolant channels, capturing thermal evolutions in regions with expected high temperatures. Additionally, line 1 represents a straight horizontal reference line, allowing for a detailed comparison of velocity and temperature variations across the core cross-section.
\smallskip

The following subsections provide a detailed discussion of the key observations derived from these simulations, with a focus on identifying the dominant thermal and flow characteristics under LOFA conditions.

\begin{figure}[H]
    \centering\includegraphics[width=0.6\linewidth]{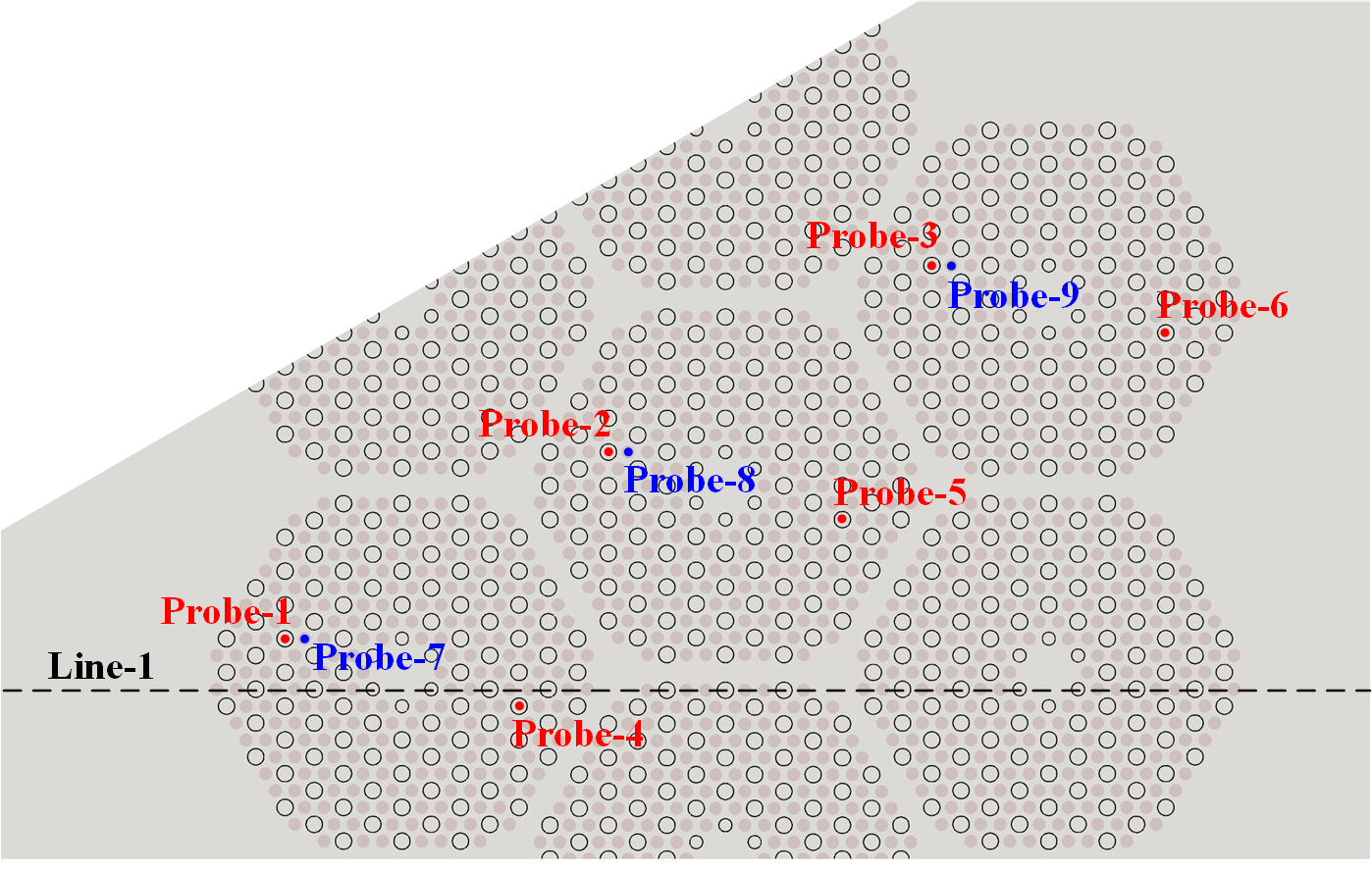}
    \caption{Probe and line locations within the cross-section of the heated section for presentation of the results.}
    \label{fig:probelinelocs}
\end{figure}

\subsubsection{Velocity distribution in coolant channels}
After the onset of LOFA, the helium circulator stops, causing a rapid reduction in coolant velocity due to loss of driving force. The forced circulation begins to decline almost immediately, and the buoyancy effects become increasingly significant. Natural circulation gradually establishes due to buoyancy forces driven by density gradients induced by temperature differences within the core and across the cooling loops. The time scale for fully developed natural circulation to establish depends on reactor design, core geometry, thermal inertia, and coolant properties. For HTGR systems, experimental and simulation-based studies generally indicate that a noticeable decrease in flow occurs within seconds to minutes after the loss of forced flow, with stable natural circulation typically forming within 10 to 30 minutes. To capture the early stage of LOFA, during which the primary heat transfer regime transitions from forced convection to natural circulation, the SubChCFD simulations were run for approximately 1000 seconds, with results extracted and analysed at t = 1000 s.

\begin{figure}[H]
    \centering\includegraphics[width=1.0\linewidth]{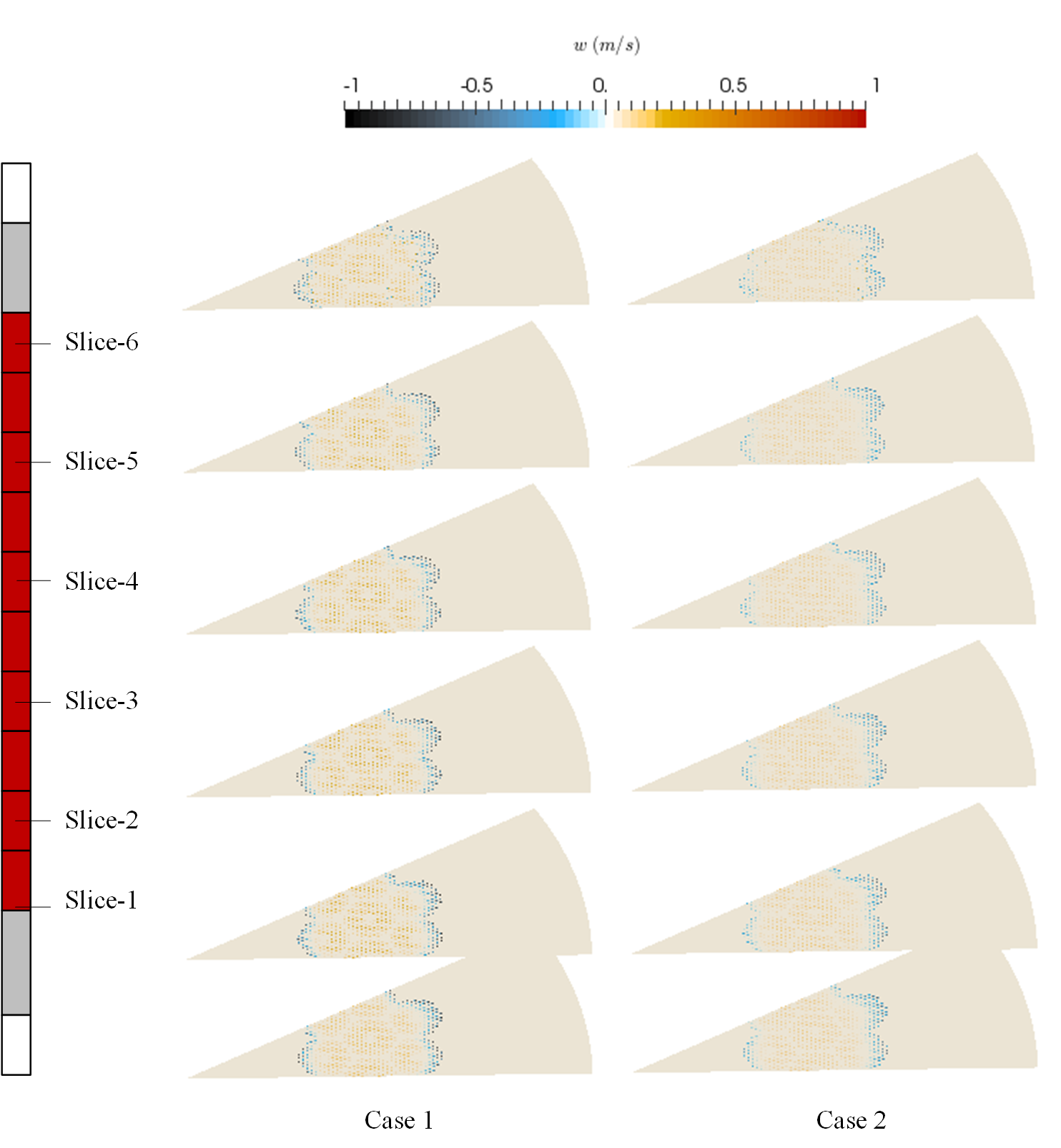}
    \caption{Velocity distributions at t = 1000 s among the coolant channels at different heights of the heated section.}
    \label{fig:wslice}
\end{figure}

The behaviour of coolant flow during LOFA conditions is a key determinant of the system’s ability to passively remove heat. Figure \ref{fig:wslice} presents the velocity distributions at t = 1000 s across multiple cross-sectional planes at different heights within the heated section. The colour scale at the top indicates the vertical velocity component ($w$) in m/s, where negative values represent downward flow, and positive values represent upward flow.
\smallskip

In both cases, a distinct circulation pattern emerges, characterised by downward flow near the periphery coolant channels of the heated fuel assemblies and upward flow concentrated in the central area. This behaviour can be attributed to the temperature-induced density difference in the coolant. Lower coolant channel temperatures are expected to appear near the periphery regions, resulting in higher coolant densities and consequently inducing downward flow due to buoyancy effects. Conversely, in the central regions, where higher temperatures are present, the coolant density is reduced, leading to the formation of buoyancy-driven upward flow. While the overall flow patterns remain similar between the two cases, slight differences are observed in the velocity magnitudes. In Case 1, the velocities are slightly higher compared to those in Case 2, indicating a stronger natural convection effect. This difference is expected, given that the decay heat generation rate is higher in Case 1, which enhances the thermal gradients within the system. In contrast, the lower decay heat generation in Case 2 produces weaker buoyancy effects, leading to a more subdued velocity field. This natural circulation pattern is a fundamental feature of passive cooling mechanisms in HTGRs, as it governs the redistribution of decay heat under LOFA conditions.

\subsubsection{Temperature distribution in the heated section}
\begin{figure}[H]
    \centering\includegraphics[width=1.0\linewidth]{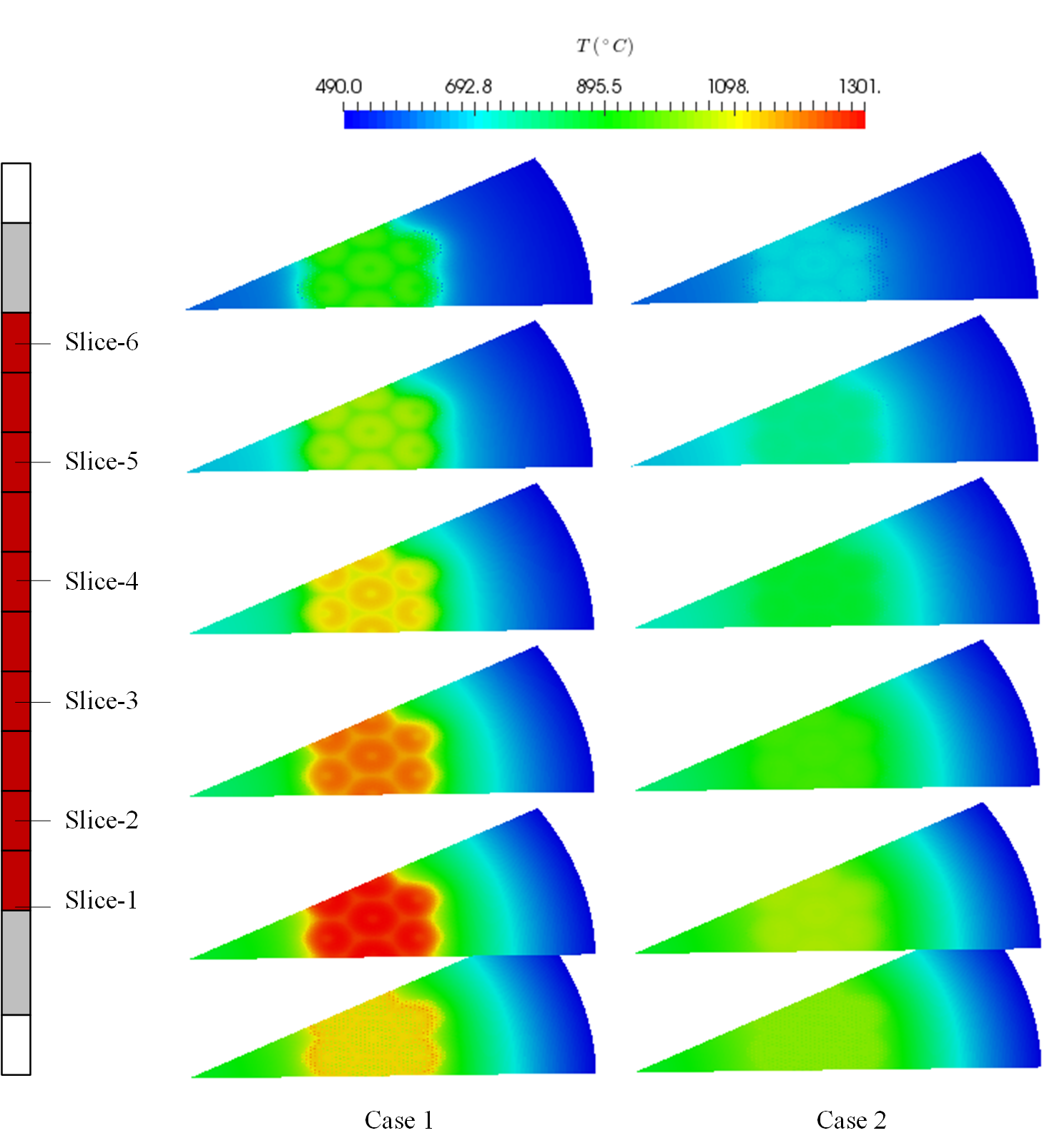}
    \caption{Temperature contours at t = 1000 s at various heights of the heated section.}
    \label{fig:Tslice}
\end{figure}

Temperature distribution is a crucial aspect in assessing the reactor’s response to a LOFA scenario. Figure \ref{fig:Tslice} presents the temperature contours at t = 1000 s at different heights of the heated section for Case 1 and Case 2. The overall trend observed in both cases is a gradual increase in temperature from the top to the bottom within the heated section, which is similar to that of steady-state operations. A notable difference between the two cases is the peak temperature observed in the core, with Case 1 exhibiting significantly higher temperatures than Case 2. This is directly attributed to the higher decay heat generation rate in Case 1, leading to a more intense thermal buildup within the fuel region. 
\smallskip

\begin{figure}[H]
    \centering\includegraphics[width=1.0\linewidth]{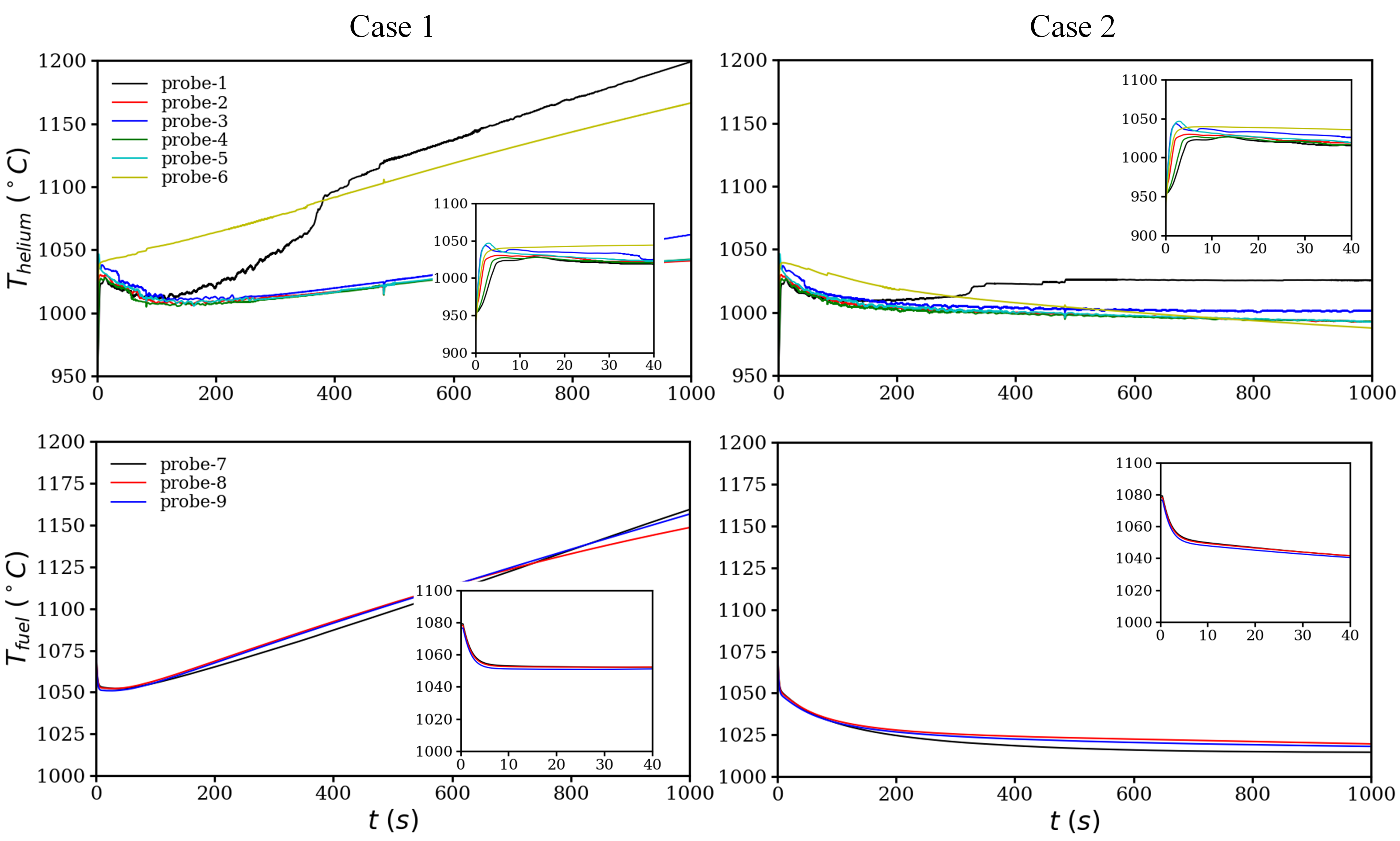}
    \caption{Time histories of helium and fuel temperature at selected cross-sectional probe locations (as shown in Figure \ref{fig:probelinelocs}) at the height of Slice 2 (as shown in Figure \ref{fig:Tslice})}.
    \label{fig:probes}
\end{figure}

Further insight into the temporal evolution of local thermal conditions is provided in Figure \ref{fig:probes}, which depicts the helium and fuel compact temperatures at multiple probe locations (see Figure \ref{fig:probelinelocs}) for the two cases. These probes are positioned at the height of slice 1 (see Figure \ref{fig:Tslice}), offering a detailed perspective on the temperature variations at fixed locations. For Case 1, the helium temperatures exhibit a three-stage pattern of evolution, characterised by an steep initial temperature rise, followed by a decrease and then a persistent temperature increase. The time period of the initial temperature rise is consistent with the rapid slowing-down of the forced flow due to pump trip, which happens within the first a few seconds after the onset of LOFA. Then, the flow undergoes a redistribution process and the temperature evolution largely depends on the flow direction change. Beyond this initial transient, the helium temperature steadily increases over time, with probe-1 and probe-6 showing the most pronounced temperature rise. A similar trend is observed in the fuel temperature evolution. The fuel undergoes a significant initial temperature drop, and the temperature stabilises within about 50 seconds thereafter. This is likely due to the initial transient heat redistribution between the solid fuel and the surrounding coolant. However, beyond this initial phase, the fuel temperature steadily rises throughout the transient period, suggesting that in Case 1, passive cooling mechanisms are insufficient to counteract the continued decay heat generation, leading to a gradual but persistent heat buildup.

In contrast, Case 2 exhibits a markedly different thermal response. The helium temperature history shows an initial temperature drop, similar to Case 1, as observed in the zoomed-in inset. However, the subsequent evolution diverges significantly. Rather than experiencing a continuous temperature rise, as seen in Case 1, the helium temperature in Case 2 gradually stabilises after approximately 300–400 seconds, forming a plateau-like behaviour. This suggests that the balance between decay heat generation and passive heat dissipation is achieved, preventing prolonged temperature escalation. The relatively small variations among different probe locations indicate a more uniform cooling effect across the core compared with Case1. A similar stabilisation effect is observed in the fuel temperature evolution. After a sharp initial decrease within the first 20–30 seconds, the fuel temperature gradually decreases and levels off at a relatively stable value. Unlike Case 1, where the fuel temperature continuously rises, Case 2 achieves a steady-state thermal condition, indicating that under lower decay heat power, passive cooling mechanisms are sufficient to prevent long-term overheating.

Overall, the comparison between the two cases highlights the crucial role of decay heat power in determining the temperature field evolution within the core. The significantly higher peak temperatures in Case 1 suggest that under high decay heat conditions, localised overheating can become a critical issue, potentially influencing material integrity and reactor safety. In contrast, the more uniform and lower-temperature distribution in Case 2 suggests that reducing decay heat effectively mitigates extreme temperature gradients. These findings emphasize the need for accurate modeling of decay heat as well as passive cooling mechanisms to ensure reliable prediction of the safe operational limits during a LOFA event.

\subsubsection{Comparisons against RANS}
\label{sec:comparisonRANS}
To further validate the predictive capabilities of SubChCFD, a comparative study was performed against RANS-based simulations, providing a benchmark for evaluating the accuracy of temperature and velocity predictions. The RANS simulations were performed based on a Pioneer Project through UK's Tier-1 HPC system ARCHER2, which aims to provide substantial computational resources to support researchers to conduct large-scale computations that require extensive processing power and parallelisation capabilities. The computational domain for the RANS model was created based on a 1/12th of the core, exactly the same as that used for the SubChCFD modelling. The $k-\omega$ SST turbulence model was used, with an all-$y^+$ wall function for the near wall treatment. To be consist with this, a computational mesh was created, consisting of approximately 900 million cells, 15 times larger than the SubChCFD mesh. Only Case 1 was considered due to the high computational cost. The simulation was conducted using 256 computational nodes (32,768 CPU cores) on ARCHER2, with approximately 120,000 node hours used, 40 times higher than those required by SubChCFD, as smaller time step sizes must be used to ensure numerical stability and accuracy.
\smallskip

\begin{figure}[H]
    \centering\includegraphics[width=1.0\linewidth]{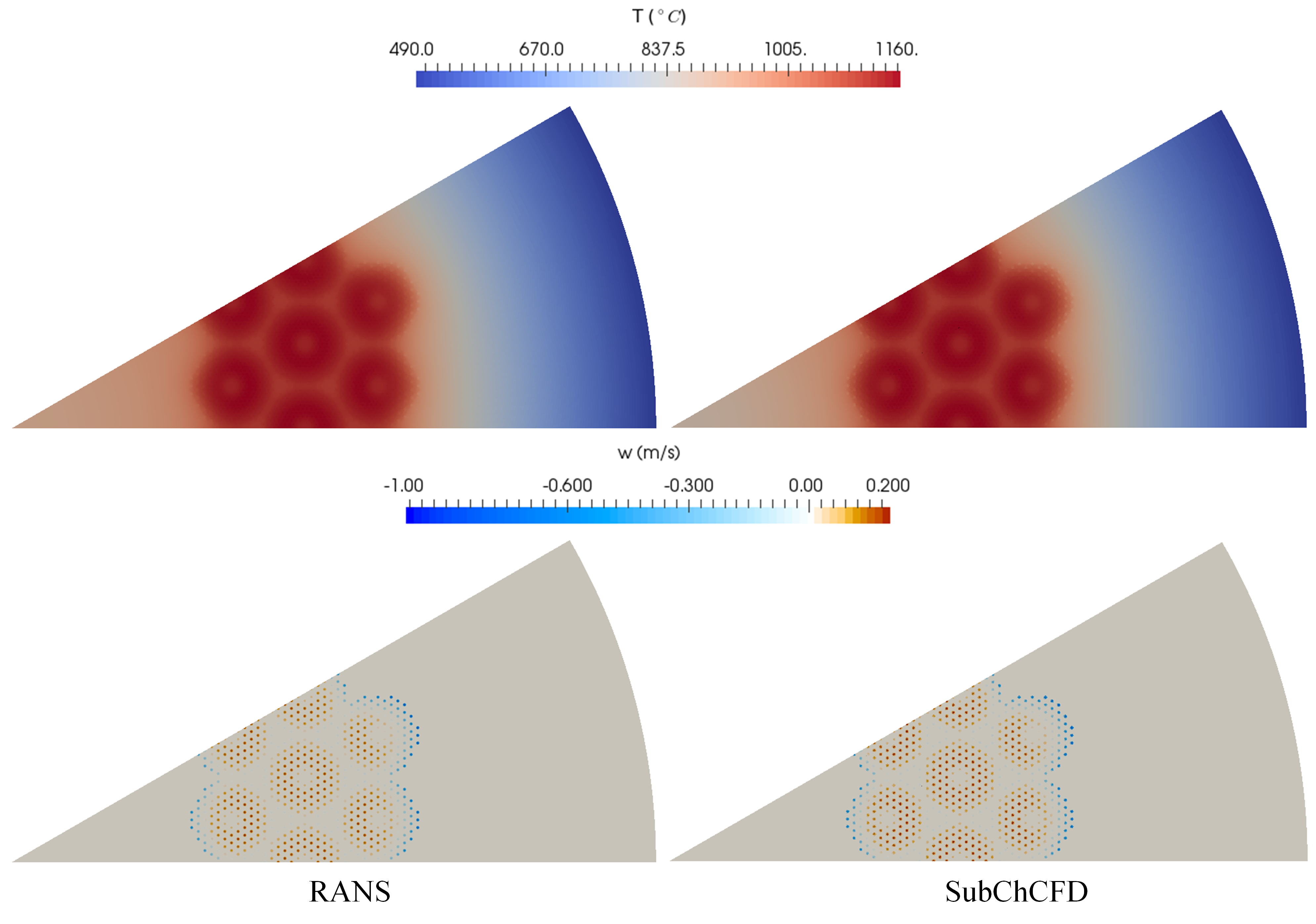}
    \caption{Comparison of temperature and velocity distributions at slice 2 at t = 503 s between SubChCFD and RANS at slice 2.}
    \label{fig:contours}
\end{figure}

The comparisons focused on temperature and velocity fields at slice 2 at t = 503 s, as shown in Figure \ref{fig:Tslice}, and further examined the distributions over line 1 of slice 2, shown in Figure \ref{fig:probelinelocs}. Figure \ref{fig:contours} presents a side-by-side comparison of temperature and velocity (the vertical component, $w$) contours between SubChCFD and RANS at slice 2 for t = 503 s. The overall temperature and velocity distribution patterns are in very good agreements between the two simulations, with both approaches capturing the key flow and thermal features of the system.

\begin{figure}[H]
    \centering
    \begin{subfigure}[t]{0.7\textwidth}
        \centering\includegraphics[width=1\linewidth]{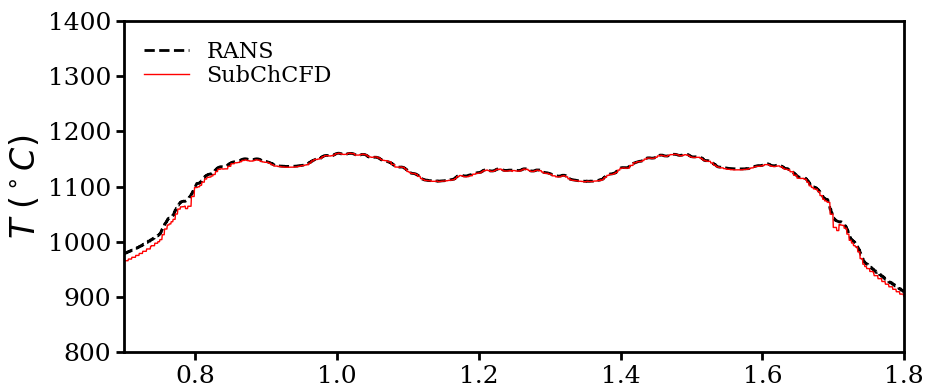}
    \end{subfigure}
    \begin{subfigure}[t]{0.7\textwidth}
        \centering\includegraphics[width=1\linewidth]{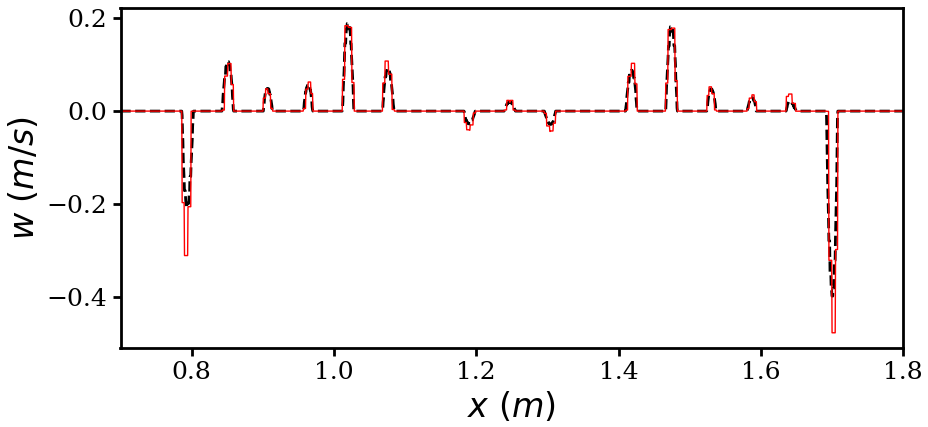}
    \end{subfigure}
    \caption{Detailed comparison of temperature and velocity variations along line 1 of slice 2 at t = 503 s.}
    \label{fig:lineplots}
\end{figure}

Detailed comparisons are provided through line plots along line 1 of slice 2, as shown in Figure \ref{fig:lineplots}. It can be observed that the temperature profiles obtained from SubChCFD exhibit good quantitative agreement with RANS, with only minor discrepancies observed near the interface between the heated fuel assemblies and the replaceable reflectors. This is consistent with a more noticeable deviation between the two simulations appears in the velocity distribution of the coolant channels in the peripheral regions of the heated fuel assemblies, where the downward coolant flow is over-predicted by SubChCFD compared with RANS. This might be due to the fact that the correction methods introduced are basically for turbulent flows, which may result in errors in cases that flow transitions from turbulence to laminar during the LOFA process. Despite these differences, it is still beyond expectation that SubChCFD produces time-accurate predictions comparable to RANS, capturing the key flow and thermal characteristics of the reactor core under LOFA conditions. Given that the computational cost of SubChCFD is significantly lower than RANS, it is doubtlessly a more efficient tool for large-scale transient studies, particularly for applications where rapid simulation turnaround and large-scale parameter exploration are needed.

\section{Conclusions}
\label{sec:conclusions}
This study has demonstrated the capability of SubChCFD as an efficient and cost-effective computational tool for simulating Loss of Flow Accidents (LOFA) in High-Temperature Gas-cooled Reactors (HTGRs). By incorporating unsteady friction model, variable property and buoyancy corrections, SubChCFD has been significantly enhanced to model the key thermal-hydraulic phenomena governing reactor behaviour during LOFA transients.
\smallskip

Through a 1/12th core model, simulations were performed to analyze two cases, each characterised by different decay heat power histories. The results revealed that temperature and velocity distributions within the reactor core are strongly influenced by the decay heat intensity, with Case 1 (higher decay heat) exhibiting stronger natural convection and higher peak temperatures compared to Case 2 (lower decay heat). Despite minor discrepancies near the interfaces between heated fuel assemblies and reflectors, SubChCFD produced temperature and velocity predictions in strong agreement with conventional Reynolds Averaged Navier-Stokes (RANS) simulations, successfully capturing the dominant natural circulation flow patterns and heat transfer mechanisms. However, the computational cost of SubChCFD is substantially lower than RANS, with at least 40 times lower consumption of computational resources as demonstrated on UK's Tier-1 HPC system ARCHER2.
\smallskip

Despite its advantages, SubChCFD still exhibits limitations, particularly in the accuracy of velocity distributions in peripheral coolant channels, where over-predictions of downward flow were observed compared to RANS. These discrepancies suggest the need for further refinements in friction and/or heat transfer models, particularly in handling laminar or transitional flows subject to influences of property variation and buoyancy. Additionally, further corrections need to be introduced to the friction models to account for the dynamic variation of wall shear stress in decelerating flows. This could effectively enhance the predictive accuracy of SubChCFD for the rapid shown-down of the flow in the initial few seconds following the onset of LOFA.
\smallskip

Overall, this study demonstrates that SubChCFD is a viable and computationally efficient alternative to convectional CFD approaches for HTGR LOFA simulations. Its ability to capture time-accurate thermal and flow behaviours at a significantly reduced computational cost makes it a powerful tool for reactor safety analysis, design optimisation, and large-scale transient assessments. With continued advancements, SubChCFD holds significant potential for expanding the predictive capabilities of reactor thermal-hydraulic modeling in future research and industrial applications.

\section*{Acknowledgements}
The present work is funded by STFC Industry Impact Fund (I2F). The authors would like to acknowledge the in-kind support received from our industry collaborator, EDF Energy R\&D UK Centre. Additionally, the authors would like to thank EPCC for providing HPC compute time through the ARCHER2 Pioneer project for passive cooling capabilities of HTGRs.
Additionally, the authors would like to thank EPCC for providing HPC compute time through the ARCHER2 Pioneer project for passive cooling capabilities of HTGRs.




\end{document}